\documentclass[journal]{IEEEtran}

\usepackage{algorithm, algorithmic}

\usepackage{subfigure}
\usepackage{textcomp}
\usepackage{stfloats}
\usepackage{url}
\usepackage{verbatim}
\usepackage{hyperref}					
\hypersetup{colorlinks,
	linkcolor=blue,%
	anchorcolor=blue,
    citecolor=blue}

\usepackage{bm}

\usepackage{cite}
\usepackage{amsmath,amssymb,amsfonts}

\usepackage{amsthm}

\theoremstyle{definition}

\newtheorem{theorem}{Theorem}

\newtheorem{lemma}{Lemma}
\newtheorem{remark}{Remark}

\usepackage{algorithmic}
\usepackage{graphicx}

\usepackage{textcomp}
\usepackage{xcolor}

\usepackage{array}

\usepackage{tabularx}

\usepackage{array}

\usepackage{booktabs}

\begin{document}
	
	
	\title{\LARGE Integrating Low-Altitude SAR Imaging into UAV Data Backhaul}
	
	\author{Zhen Du,~\IEEEmembership{Member,~IEEE,} Fan Liu,~\IEEEmembership{Senior Member,~IEEE}, Jie Yang,~\IEEEmembership{Member,~IEEE}, Yuanhao Cui,~\IEEEmembership{Member,~IEEE,} Rui Wang,~\IEEEmembership{Senior Member,~IEEE,} Zenghui Zhang,~\IEEEmembership{Senior Member,~IEEE} and~Shi Jin,~\IEEEmembership{Fellow,~IEEE}
		\thanks{(\textit{Corresponding author: Fan Liu})}
		\thanks{Zhen Du is with the School of Electronic and Information Engineering, Nanjing University of Information Science and Technology, Nanjing 210044, China (email: duzhen@nuist.edu.cn).
		}	
		\thanks{Fan Liu and Shi Jin are with the National Mobile Communications Research Laboratory, Southeast University, Nanjing 210096, China (e-mail: fan.liu; jinshi@seu.edu.cn).
		}
		\thanks{Jie Yang is with the Key Laboratory of Measurement and Control of Complex Systems of Engineering, Ministry of Education, and the Frontiers Science Center for Mobile Information Communication and Security, Southeast University, Nanjing 210096, China (e-mail: yangjie@seu.edu.cn).
		}
		\thanks{Yuanhao Cui is with the School of Information and Electronic Engineering, Beijing University of Posts and Telecommunications, Beijing 100876, China (e-mail: yuanhao.cui@bupt.edu.cn).}
		\thanks{Rui Wang is with the Department of Electrical and Electronic Engineering, Southern University of Science	and Technology (SUSTech), Shenzhen 518055, China (e-mails: wang.r@sustech.edu.cn)}
		\thanks{Zenghui Zhang  is with the Shanghai Key Laboratory of Intelligent Sensing and Recognition, Shanghai Jiao Tong University, Shanghai 200240, China (e-mail: zenghui.zhang@sjtu.edu.cn).
		}
	}

	\maketitle
	
	\begin{abstract}
		Synthetic aperture radar (SAR) on unmanned aerial vehicles (UAVs) enables high-resolution sensing in low-altitude wireless networks, while requiring reliable uplink data backhaul to ground base stations under dynamic channel conditions. Conventional orthogonal frequency division multiplexing (OFDM)-based SAR systems rely on pilot or deterministic signaling, which occupies only a small fraction of the available time-frequency (TF) resources and limits imaging performance. This paper develops a data-aided OFDM-SAR imaging framework that reuses uplink communication data symbols for sensing, thereby exploiting the dominant TF resources of the UAV backhaul link. However, the randomness of data symbols disrupts the coherent structure required for SAR imaging, especially in highly dynamic channels with strong TF coupling, leading to severe degradation in range-Doppler focusing. To address this issue, we establish a unified TF domain filtering framework to suppress data-induced randomness and recover an equivalent deterministic imaging channel. Within this framework, reciprocal, matched, and Wiener filtering are interpreted under a common formulation, enabling a systematic characterization of their impact on imaging performance. A normalized mean square error (NMSE) metric of a reference point target’s profile is further adopted to quantify the joint effects of randomness-induced distortion and noise amplification. Simulation results based on 5G NR parameters show that the proposed data-aided scheme significantly outperforms pilot-only approaches by leveraging uplink data resources, demonstrating that effective TF-domain filtering is essential to ensure high-resolution imaging in dynamic UAV channels.
	\end{abstract}


	\section{Introduction}\label{sec1}
	\IEEEPARstart{W}{ith} the rapid advances of the low-altitude economy, unmanned aerial vehicles (UAVs) are increasingly deployed as key platforms for civilian and industrial missions such as urban inspection, logistics, infrastructure monitoring, and environmental observation. Benefiting from their mobility, low cost, and flexible deployment in low-altitude wireless networks (LAWNs), UAVs provide an efficient means for large-scale data collection and real-time situational awareness \cite{wu2025low,wu2025toward,song2025overview,lyu2022joint}. In particular, when equipped with synthetic aperture radar (SAR) sensors \cite{cumming2005digital,richards2005fundamentals,wang2013mimo,zhang2024optical}, UAVs are capable of providing high-resolution and all-weather imaging that is robust to illumination and atmospheric variations. Therefore, UAV-SAR in LAWNs enables accurate environmental perception and mapping, which are essential for urban management, intelligent transportation, and public security. 
	
	However, in many time-critical missions, UAV-SAR imaging must satisfy harsh real-time requirements. For instance, in urban traffic monitoring and moving target tracking, continuous imaging and low-latency data backhaul are crucial for adaptive control and rapid decision-making \cite{du2025toward}. This imposes dual challenges on UAV-SAR systems: the need for precise, high-resolution imaging, and the requirement for high-throughput, low-latency data transmission.
	To this end, communication and sensing modules on current UAV platforms are typically implemented separately. The sensing module integrates various sensors such as radar, LiDAR, and cameras, which occupy significant space and weight while introducing considerable power consumption, thereby limiting the UAV’s endurance. 
	To address this issue, integrating radar sensing and communications into a unified module is a promising solution. Current literature exploit a unified integrated sensing and communications (ISAC) signaling strategies, mainly focusing on sensing-centric \cite{wang2019first,chatzitheodoridi2022cooperative,huang2025design} and joint designs \cite{yang2022waveform}.
	In contrast, communication-centric designs \cite{sturm2011waveform,kumari2017ieee,cui2021integrating,wang2026clutter,lu2025sensing}, which has received comparatively less attention in SAR imaging, may offer a potentially more practical and cost-effective solution. By reusing the existing communication hardware for imaging purposes, this approach not only improves spectrum efficiency but also helps mitigate mutual interference between the imaging and communication functions.
	In this spirit, joint orthogonal frequency-division multiplexing (OFDM) SAR imaging and data transmission has emerged as a promising approach. By leveraging the OFDM waveform, the system can simultaneously achieve high-quality SAR imaging and communications, thus formulating a new ISAC paradigm \cite{liu2022integrated} in LAWNs.

	Standardized OFDM has been widely employed as the default waveform in cellular communications\cite{hwang2008ofdm} and WiFi \cite{niu2022rethinking}, due to its high spectral efficiency and robustness against multipath fading. In addition, the OFDM waveform has been increasingly employed in radar systems for target detection, estimation and tracking tasks \cite{berger2010signal,shi2017power,bicua2016generalized,sen2010adaptive,du2020distributed}, owing to its flexible spectrum design and thumbtack-shaped ambiguity function. In particular, its linear temporal-frequency (TF) structure allows straightforward delay-Doppler information extraction using 2D-fast Fourier transform (FFT), while the embedding of cyclic prefix (CP) provides robustness against frequency-selective fading \cite{du2025prob}. 
	In addition to the sensing functionalities above, the radar imaging capability of exploiting OFDM signals, has also been frequently reported in the last decade \cite{garmatyuk2003ultra,garmatyuk2011adaptive,zhang2014ofdm,zhang2014irci,yu2018irci,wang2019first}. The motivation also benefits from the exploitation of multicarrier structure and the CP, which can readily synthesize a large bandwidth signaling, and mitigate the inter-range cell interference (IRCI) induced by the inter-symbol interference (ISI).

	In spite of this, aforementioned works replace the random data of OFDM communications signals with time-repeated deterministic/pseudo noise-like weighting coefficients \cite{garmatyuk2003ultra,garmatyuk2011adaptive,zhang2014ofdm,zhang2014irci,yu2018irci}, or multiplex sensing and communications in different resource blocks \cite{wang2019first}. As such, the data transmission capability is significantly sacrificed. 
	Notably, the above sensing signaling may be treated as pilots in the context of 5G New Radio (NR) frame structure, such as the sounding reference signals (SRS) \cite{dahlman20205g} in the uplink channel. However, SRS occupies only a small fraction of TF resources, resulting in limited bandwidth, pulse duration, and energy accumulation, which in turn degrades imaging resolution and accuracy. Moreover, the low pulse repetition frequency (PRF) of pilot transmissions may lead to Doppler aliasing due to reduced azimuth sampling density, thereby impairing the azimuth focusing performance. In addition, SRS has a discontinuous comb structure across subcarriers \cite{etsi38138}, which leads to periodic peaks in the time domain and thus deteriorate the imaging performance.
	On top of these factors, most downlink signals from ground base stations to UAVs merely convey control instructions, while the UAV must upload large volumes of imaging data, leading to a much higher uplink backhaul throughput. To comprehensively balance imaging and communication qualities, together with computational complexities, the communication-centric signaling design \cite{sturm2011waveform,kumari2017ieee,cui2021integrating} maintaining compatibility with the existing 5G NR protocol \cite{dahlman20205g} is thus envisioned as a more economically viable paradigm with data symbols for imaging.
	
	Nevertheless, data-aided imaging triggers a new challenge: the randomness must be effectively mitigated to restore the orthogonality between the range and azimuth steering vectors through proper TF-domain filtering schemes \cite{du2024reshaping}, thereby ensuring compatibility with conventional 2D Fourier-based imaging algorithms such as range-Doppler (RD) and Omega-$K$ methods \cite{cumming2005digital}.
	To be specific, data symbols are typically drawn from quadrature amplitude modulation (QAM) constellations, which are differ both in time and frequency dimensions \cite{du2024reshaping}. Consequently, the range and azimuth profiles may be simultaneously jeopardized by the signaling randomness, leading to compromised range-azimuth compression performance. 
	To mitigate this effect, data-dependent filtering can be performed in the TF domain, such as matched filtering \cite{kay1993fundamentals}, element wise division \cite{sturm2011waveform,zhang2014ofdm} or linear minimum mean squared error (LMMSE) filtering \cite{du2025prob, keskin2024fundamental}. However, a quantitative characterization of imaging performance under data-aided scheme remains unclear, particularly in terms of metrics that can capture the impact of signaling randomness and filtering design.

	Based upon the above reasoning, this article aims to integrate the OFDM-SAR imaging into UAV data backhaul in LAWNs, explore the superiority of introducing data-aided imaging relative to pilot-only imaging, and reveal the relationship between the imaging quality and the ISAC signaling randomness.
	For clarity, the major contributions of this paper are summarized as follows.
	
	\begin{itemize}
		\item Unlike conventional OFDM-SAR designs inspired by 5G NR, which effectively exploit only pilot signals (e.g., SRS) for sensing, we establish a data-aided imaging model that explicitly incorporates data symbols into both range and azimuth dimensions. Since the inherent randomness of communication data can noticeably distort the echo statistics and degrade imaging quality, the proposed TF-domain filters are designed to suppress such randomness, thereby regularizing the echo structure and enabling it compatible with the standard RD algorithm.
		\item Furthermore, we establish a quantitative relationship between ISAC signaling randomness and SAR imaging performance by adopting the normalized mean square error (NMSE) of a reference point target. Unlike conventional metrics that evaluate physical attributes in isolation, the NMSE provides a supplementary perspective that unifies the joint effects of signaling-induced sidelobes, peak energy loss, and noise amplification into a single indicator. This framework allows for a more direct characterization of how discrete QAM modulation impacts the overall reconstruction in data-aided SAR systems.
	\end{itemize}

	The remainder of this paper is organized as follows. Sec. \ref{sec2} introduces the system model. Sec. \ref{sec3} develops the OFDM-SAR imaging approach. Sec. \ref{sec4} elaborates on the sensing criteria. Sec. \ref{sec5} provides simulation results to validate the theoretical analysis. Finally, Sec. \ref{sec6} concludes this article.

	\textit{Notation:} Throughout the paper, $\mathbf{A}$, $\mathbf{a}$, and $a$ denote a matrix, vector, and scalar, respectively; $\vert a\vert$, $\vert \mathbf{A}\vert$ and $\frac{\mathbf{B}}{\mathbf{A}}$ represent the modulus of $a$, the element-wise modulus of $\mathbf{A}$ and the element-wise division of $\mathbf{B}$ by $\mathbf{A}$, respectively;
	$\mathbb{E}(\cdot)$, $tr(\cdot)$, $\Vert\cdot\Vert_F$, $(\cdot)^{T}$, $(\cdot)^{*}$, $(\cdot)^{H}$, $\odot$ and $\mathbf{I}$ denote the expectation, trace, Frobenius norm, transpose, conjugate, Hermitian, Hadamard (element-wise) product and identity matrix, respectively; A complex Gaussian distribution with mean $\mu$ and variance $\sigma^2$ is represented as $\mathcal{CN}(\mu,\sigma^2)$; Finally, $\mathrm{sinc}(x)=\frac{\sin(\pi x)}{\pi x}$.

	\begin{figure}[!t]
		\centering
		\includegraphics[width=0.8\linewidth]{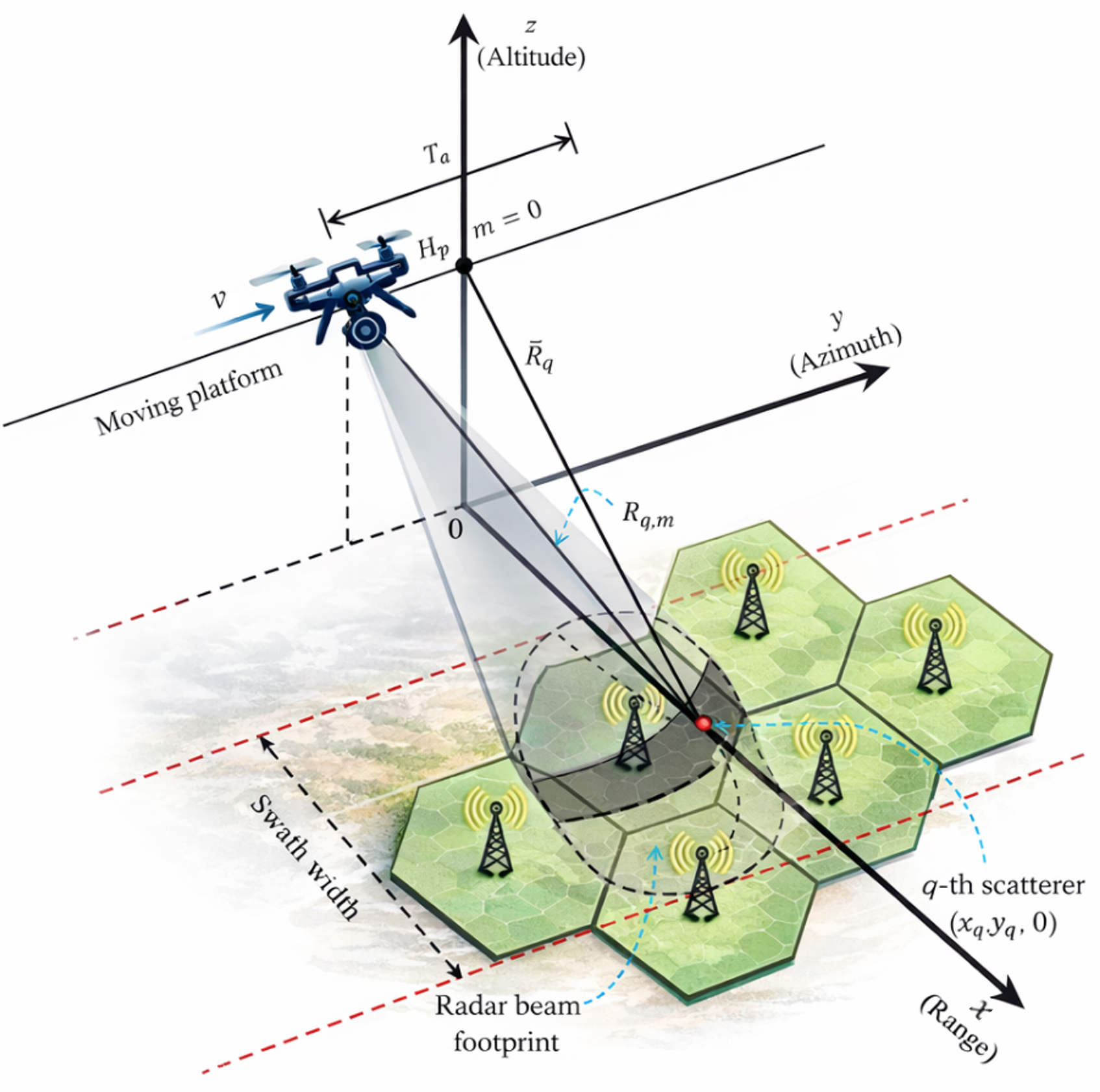}
		\caption{Joint low-altitude OFDM-SAR imaging and data backhaul.}
		\label{figure0}
	\end{figure}

	\section{System Model}\label{sec2}
	\subsection{SAR Geometry}
	As depicted in Fig.~\ref{figure0}, we consider a joint monostatic broadside stripmap UAV-SAR imaging embedded in continuous data backhaul scheme with cellular base stations in LAWNs.
	The UAV moves along the $y$-axis, with a height of $H_p$ and a constant velocity of $v$.
	To guarantee uninterrupted communications, the UAV's coverage footprint must exceed the radius of a base station, ensuring at least one base station remains in line-of-sight of SAR beampattern at all times. Consequently, a quantitatively feasibility analysis is required.
	
	Let us commence by analyzing the azimuth beamwidth and the elevation beamwidth, expressed as \cite{cumming2005digital}
		\begin{align}
			\theta_\mathrm{az} \approx  \frac{0.886\lambda}{D_a} \ \text{and} \
			\theta_\mathrm{el} \approx  \frac{0.886\lambda}{D_e},
		\end{align}
	where $\lambda$ is the wavelength, $D_\mathrm{al}$ and $D_\mathrm{el}$ represent  azimuth and elevation antenna apertures, respectively. With a typical height $H_p$ from $300$ to $3000$ meters in LAWNs, the center of SAR elevation beamwidth defines the beam’s pointing elevation angle $\theta_c$. Accordingly, the azimuth coverage length $L_\mathrm{az}$ and the elevation coverage length $L_\mathrm{el}$ can be approximated as
	\begin{align}
		&L_\mathrm{az} \approx  \frac{2H_p}{\cos\left(\theta_c\right)}\tan\left(\frac{\theta_\mathrm{az}}{2}\right),
		\\
		L_\mathrm{el} \approx   H_p&\left(\tan\left(\theta_c+\frac{\theta_\mathrm{el}}{2}\right)-\tan\left(\theta_c-\frac{\theta_\mathrm{el}}{2}\right)\right).
	\end{align}

	As clarified above, in order to guarantee the real-time data backhaul, $L_\mathrm{az}$ and $L_\mathrm{el}$ must be larger than the cellular coverage. 
	For example, we consider $3.5$ GHz carrier frequency in sub-6 GHz band, azimuth and elevation antenna apertures as $D_\mathrm{az}=D_\mathrm{el}=0.1$ m  and a central elevation angle $\theta_c=\frac{\pi}{4}$. 
	At present, most commercial low-altitude UAV SAR platforms fly at around $H_p=1000$ m to balance spatial coverage and echoes' magnitude.
	In this case, the corresponding ground coverage lengths are approximately $L_\mathrm{az}\approx 564$ m and $L_\mathrm{el}\approx 1899$ m, which obviously exceed the coverage of a 5G cellular cell. 
	Based on these parameter analysis, the practical feasibility of integrating OFDM-SAR imaging into UAV data backhaul in LAWNs is geometrically convinced\footnote{Low-altitude UAV can also employ a dual-beam mode: an imaging beam is utilized for ground sensing, while data backhaul is conducted through a dedicated directional beam pointing toward a fixed base station. This involves ISAC beam management issues \cite{du2025toward}, which are beyond the scope of this paper.}.

	\subsection{Transmit ISAC Signal Model}	
	The UAV platform transmits unified ISAC OFDM signals for dual functionalities of imaging and data backhaul tasks simultaneously. The signals consisting of $N$ subcarriers and $M$ symbols, and occupying a bandwidth of $B_r$ Hz, a symbol duration of $T_p$ seconds, and a synthetic aperture azimuth time of $T_a$ seconds, is given by
	\begin{equation}\label{eq1}
		\begin{aligned}
			x(t) = \frac{1}{\sqrt{M}}\sum\nolimits^{M-1}_{m=0}x_m(t-mT_{\mathrm{sym}}),
		\end{aligned}
	\end{equation}
	where
	\begin{equation}\label{eq2}
		\begin{aligned}
			x_m(t) = \frac{1}{\sqrt{N}}\sum\nolimits^{N-1}_{n=0} s_{n,m} e^{j2\pi n \Delta ft} \mathrm{rect}\left( \frac{t-mT_{\mathrm{sym}}}{T_{\mathrm{sym}}} \right).
		\end{aligned}
	\end{equation}
	In (\ref{eq2}), $s_{n,m}$ denotes the transmitted frequency-domain data, drawn from a finite alphabet such as 256-QAM constellation.
	In addition, $\Delta f = B_r/N=1/T_p$ represents the subcarrier interval in the frequency domain, and $\mathrm{rect}(t)$ represents the rectangle window, equal to $1$ for $0\leq t\leq 1$, and zero otherwise\footnote{Note that the practical SAR has a sinc-like azimuth envelope \cite{cumming2005digital,zhang2014ofdm}. For the theoretical convenience, we use an ideal rectangle envelope instead.}. Additionally, $T_{\mathrm{sym}}=T_p+T_{cp}$, where $T_{cp}$ denotes the length of CP. In order to eliminate the ISI and the inter-carrier interference (ICI) of received sensing signals, $T_{cp}$ must be larger than round-trip delay of the farthest target/path \cite{du2025prob,keskin2024fundamental,keskin2021mimo}. The total azimuth time is thus $T_a=MT_\mathrm{sym}$.
	Note that the power of a given codebook is normalized, i.e., $\mathbb{E}\{|s_{n,m}|^2\}=1$ for $\forall n,m$. This leads to the fact that the total average transmit energy of $M$ symbols is also normalized as $\mathbb{E}\{|x(t)|^2\} = \frac{1}{NM}\mathbb{E}\left\{\sum_{n,m}|s_{n,m}|^2\right\}=1$.

	
	
	
	Finally, through the up converter, the ISAC signal is transmitted at the RF frequency as
	\begin{equation}\label{eq3}
		\begin{aligned}
			\Re\{x(t)\exp\left(j2\pi f_ct\right)\},
		\end{aligned}
	\end{equation}
	where $f_c$ denotes the carrier frequency.

	\subsection{SAR Echo Model}

	We assume that the received OFDM echo are reflected by $Q$ resolvable scatterers/pixels, where the $q$th one is positioned on the grid of $(x_q, y_q)$. In practical low-altitude UAV SAR scenarios, the wireless channel is inherently time-varying due to platform motion, introducing Doppler-induced phase variations across the coherent processing interval. Then the echo in the TF domain can be formulated as\footnote{Under typical 5G-NR configurations, the channel dynamics-induced ICI effect in a moving-platform SAR system remains limited. Specifically, the maximum normalized Doppler shift $\eta=\frac{2vf_c}{c\Delta f} \approx 0.04$ for $f_c = 3.5$ GHz, $\Delta f = 30$ kHz, and $v = 50$ m/s. Notably, this corresponds to the worst case, while in practical side-looking SAR geometries the effective Doppler is much smaller. Therefore, the induced ICI is generally weak as $\eta< 0.1$ \cite{sturm2011waveform} and can be reasonably neglected in subsequent modeling.}
	\begin{equation}\label{equ1}
		\begin{aligned}
			y_{n,m} = & \sum\nolimits^Q_{q=1}\bar{\alpha}_q s_{n,m}e^{-j2\pi (f_c+n\Delta f)\frac{2R_{q,m}}{c}}+z_{n,m},
		\end{aligned}
	\end{equation}
	where $c$, $\bar{\alpha}_q$, $R_{q,m}$ and $z_{n,m}$ represent the speed of light, the radar cross section (RCS) coefficient caused by the $q$th scatterer within the radar beam footprint, the instantaneous slant range between the UAV and the $q$th scatterer, and the additive white Gaussian noise, respectively. For analytical convenience, it is assumed that $\bar{\alpha}_q\sim\mathcal{CN}(0,\sigma^2_{\alpha_q})$ and $z_{n,m}\sim\mathcal{CN}(0,\sigma^2)$.
	
	Next, we use the first-order slant range approximation as 
	\begin{equation}\label{equ2}
		\begin{aligned}
			R_{q,m}= & \sqrt{x^2_q+H^2_p+(vmT_{\mathrm{sym}}-y_q)^2}
			\\ \approx & \bar{R}_q+ \underbrace{\frac{(vmT_{\mathrm{sym}}-y_q)^2}{2\bar{R}_q}}_{\Delta R_{q,m}},
		\end{aligned}
	\end{equation}
	where $\bar{R}_q=\sqrt{x^2_q+H^2_p}$. Note that the approximation holds when the azimuth synthetic aperture length $vT_a$ is much smaller than the slant range $\bar{R}_q$ and the radar operates under a small squint angle \cite{cumming2005digital}. In LAWNs, $vT_a$ of the UAV can reach on the order of hundreds of meters, while its altitude can be on the order of several thousands of meters. In addition, the small-squint-angle condition is naturally satisfied in broadside mode. Therefore, the approximation in (\ref{equ2}) holds.
	
	Then substituting (\ref{equ2}) into (\ref{equ1}) and rearranging terms yields
	\begin{equation}\label{equ14}
		\begin{aligned}
			y_{n,m} = \sum^Q_{q=1}\alpha_q s_{n,m} & \underbrace{e^{-j\frac{4\pi}{c}n\Delta f \Delta R_{q,m}}}_{\text{TF Coupling: RCM}} \underbrace{e^{-j\frac{4\pi}{c}n\Delta f \bar{R}_q}}_{\text{Range Term}} \\ & \times \underbrace{e^{-j\frac{4\pi}{c} f_c\Delta R_{q,m}}}_{\text{Azimuth Term}} +z_{n,m},
		\end{aligned}
	\end{equation}
	where $\alpha_q=\bar{\alpha}_q e^{-j\frac{4\pi}{c} f_c \bar{R}_q}$. In addition to separated range and azimuth terms, the range-cell migration (RCM) term makes conventional 2D-FFT approach in \cite{du2025prob,keskin2024fundamental} difficult to achieve independent range-Doppler focusing. While the CP effectively mitigates ISI, the high-mobility trajectory of the UAV introduces a time-variant phase coupling across the TF grid. This RCM effect is essentially the manifestation of time-variant frequency selectivity, where the channel response evolves across both subcarriers and symbols. The subsequent TF-domain filtering framework is designed to embrace these dynamics by regularizing the random waveform, thereby restoring the coherence required for high-resolution imaging.


	To proceed, we may reformulate (\ref{equ14}) with a more compact matrix form as
	\begin{equation}\label{equ15}
		\begin{aligned}
			\mathbf{Y} = \mathbf{H}\odot\mathbf{S}+\mathbf{Z},
		\end{aligned}
	\end{equation}
	where $\mathbf{S}\in\mathbb{C}^{N\times M}$ and $\mathbf{Z}\in\mathbb{C}^{N\times M}$ represent the random symbol and noise matrices, respectively, and $\mathbf{H}\in\mathbb{C}^{N\times M}$ denotes the imaging channel matrix, expressed as
	\begin{equation}\label{equ16}
		\begin{aligned}
			\mathbf{H} = \sum\nolimits^Q_{q=1} \alpha_q \underbrace{\mathbf{b}\mathbf{c}^H\odot\mathbf{E}}_{\mathbf{H}_q}.
		\end{aligned}
	\end{equation}
	Here, $\mathbf{b}\in \mathbb{C}^{N\times 1}$ with the $n$th element as $[\mathbf{b}]_n = e^{-j\frac{4\pi}{c}n\Delta f \bar{R}_q}$, $\mathbf{c}\in \mathbb{C}^{M\times 1}$ with the $m$th element as $[\mathbf{c}]_m=e^{-j\frac{4\pi}{c} f_c\Delta R_{q,m}}$, and $\mathbf{E}\in \mathbb{C}^{N\times M}$ with the $(n,m)$th element as $[\mathbf{E}]_{n,m}=e^{-j\frac{4\pi}{c}n\Delta f \Delta R_{q,m}}$.

	\begin{remark}
		The reconstruction of OFDM-SAR imaging can be expressed as a linear model by reformulating (\ref{equ15}) as
		\begin{equation}\label{parameter}
			\begin{aligned}
				{\mathbf{y}} = \tilde{\mathbf{H}}\bm{\alpha}+{\mathbf{z}},
			\end{aligned}
		\end{equation}
		where ${\mathbf{y}}=\mathrm{vec}(\mathbf{Y})$, $\tilde{\mathbf{H}}=\left[\mathrm{vec}(\mathbf{H}_1\odot\mathbf{S}),\cdots,\mathrm{vec}(\mathbf{H}_Q\odot\mathbf{S})\right]$, $\bm{\alpha}=[\alpha_1,\cdots,\alpha_Q]^T$, and $\mathbf{z}=\mathrm{vec}(\mathbf{Z})$.
		Since $\mathbf{z}$ is Gaussian and satisfies $\mathbb{E}\{\mathbf{z}\}=\mathbf{0}$ and $\mathbf{R}_z=\mathbb{E}\{\mathbf{z}\mathbf{z}^H\}=\sigma^2\mathbf{I}$, 
		from a parameter estimation perspective, SAR imaging can thus be interpreted as the least-squared (LS) estimation of $\bm{\alpha}$ \cite{zhang2014ofdm,zhang2014irci}
		\begin{equation}
			\begin{aligned}
				\hat{\bm{\alpha}} = \left(\tilde{\mathbf{H}}^H\tilde{\mathbf{H}}\right)^{-1}\tilde{\mathbf{H}}^H\mathbf{y}.
			\end{aligned}
		\end{equation}
		
		Since the number of pixels $Q$ is typically enormous, $\tilde{\mathbf{H}} \in \mathbb{C}^{NM \times Q}$ becomes too large to store, making direct LS inversion computationally intractable. Instead, SAR commonly resorts to FFT-based methods, e.g., range-Doppler or Omega-K algorithms \cite{cumming2005digital}, which are typically implemented via coherent phase-history processing, such as range compression, azimuth focusing, and Doppler-domain processing.
		Thanks to the inherent Fourier structure of $\tilde{\mathbf{H}}$ to avoid the direct inversion of a prohibitively large matrix, these widely used algorithms can be interpreted as computationally efficient approximations to the LS solution, which achieve near-optimal focusing performance with significantly reduced complexity.
		This connection provides a useful bridge between the proposed OFDM-based formulation and conventional SAR imaging mechanisms, which is further clarified in Sec.~\ref{sec3}.
	\end{remark}

	


	\section{SAR Imaging Approach}\label{sec3}
\begin{figure}[!t]
		\centering
			\includegraphics[width=0.95\linewidth]{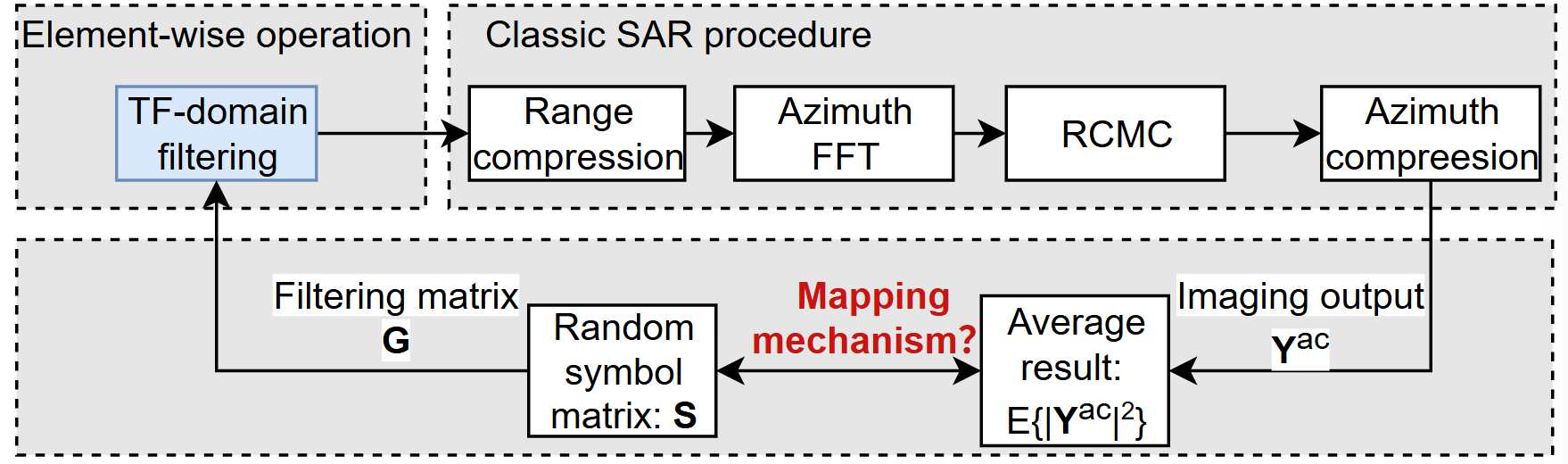}
		\caption{Technical roadmap.}
		\label{f0}
	\end{figure}

	In this section, we elaborate on the procedure of integrating SAR imaging into UAV data backhaul, where the technical roadmap is shown in Fig.~\ref{f0}. There are two key differences compared to the conventional RD algorithm. First, we propose to exploit several TF-domain filtering schemes to allieviate the impact of data randomness. In this work, the TF-domain filtering is not introduced as a new estimator, but as a means to establish the mapping from random communication data symbols to SAR imaging performance. This mapping enables a systematic analysis of how signal randomness propagates through the imaging chain and affects the final image quality.
	Second, the azimuth signal is not a strict quadratic-phase signal, since the random data imposes amplitude modulation on it, and the approximation of its Doppler spectrum requires further discussion.
	
	
	\subsection{{TF Filtering}} 
	Different from conventional SAR imaging based on chirp waveforms, SAR using OFDM communication signals may suffer from severe imaging degradation due to the modulation randomness and the employed TF-domain filtering schemes. 
	To mitigate this effect, the first step is to compensate for the discrete symbols drawn from a given QAM constellation.

	To proceed, we now reformulate (\ref{equ15}) as
	\begin{equation}\label{equ17}
		\begin{aligned}
			\mathbf{y} = \mathbf{A}\mathbf{h}+\mathbf{z},
		\end{aligned}
	\end{equation}
	where $\mathbf{A}=\mathrm{diag}(\mathrm{vec}(\mathbf{S}))$, and $\mathbf{h}=\mathrm{vec}(\mathbf{H})$ which involves $\bm{\alpha}$. Then we arrive at $\mathbb{E}\{\mathbf{h}\}=\mathbf{0}$ and $\mathbf{R}_h=\mathbb{E}\{\mathbf{h}\mathbf{h}^H\}=\sum^Q_{q=1}\sigma^2_{\alpha_q}\mathbf{I}$, which can be similarly proved by referring to \cite[Lemma 1]{keskin2024fundamental}.
	
	
	\begin{remark}\label{remark2}
		By rewriting the model as (\ref{equ17}), we shift the focus to the channel (observation) space, where $\mathbf{h}$ represents the equivalent imaging channel. The diagonal structure of $\mathbf{A}$ decouples the observations, enabling efficient $NM$ parallel estimation. Notably, from a parameter-space perspective like that in (\ref{parameter}), estimating $\mathbf{h}$ implicitly contains information of $\bm{\alpha}$.
		Consequently, this formulation provides a natural two-stage estimation strategy:
		1) channel-space estimation: estimate $\mathbf{h}$ independently for each subchannel;
		2) parameter-space inversion: recover the physical scatterer parameters $\bm{\alpha}$ from $\mathbf{h}$ in accordance with their linear relationship as $\mathbf{h}=\mathrm{vec}(\mathbf{H})=\mathrm{vec}\left(\sum_q \alpha_q\mathbf{H}_q\right)=\sum_q \alpha_q\mathrm{vec}\left(\mathbf{H}_q\right)$, through 2D-FFT-based algorithms.
	\end{remark}
	
	Our current aim is to estimate $\mathbf{h}$, formalized as $\hat{\mathbf{h}}=\mathbf{G}\mathbf{y}$, where $\mathbf{G}$ represents the TF-domain filtering matrix. 
	We establish a unified TF-domain estimation framework for OFDM-SAR imaging under random data modulation, where TF-domain filtering is interpreted as estimating the underlying imaging channel rather than merely performing symbol compensation. Unlike conventional OFDM-SAR systems that employ deterministic and periodically repeated waveforms across pulses, the considered framework utilizes OFDM communication signals, where data symbols vary randomly, introducing random modulation effects into the imaging process.
	
	Within this framework, classical strategies such as reciprocal filtering (RF), matched filtering (MF), and Wiener filtering (WF), are revisited as estimators under different optimality criteria. This unified perspective enables a systematic analysis of how different filtering strategies affect imaging quality, and provides new insights into the mapping mechanism between signal randomness and SAR imaging, which is not captured in conventional systems with deterministic waveforms.

	\subsubsection{Reciprocal Filtering (RF)}
	RF is designed based on the least-squared estimation, with its objective to minimize the squared error as
	\begin{equation}
		\begin{aligned}
			\hat{\mathbf{h}}_\mathrm{RF}=\arg\min_{\mathbf{h}} \ \Vert\mathbf{y}-\mathbf{A}\mathbf{h}\Vert^2_F,
		\end{aligned}
	\end{equation}	
	which yields the solution as
	\begin{equation}\label{eq17}
		\begin{aligned}
			\hat{\mathbf{h}}_\mathrm{RF}=\underbrace{\left(\mathbf{A}^H\mathbf{A}\right)^{-1}\mathbf{A}^H}_{\mathbf{G}_\mathrm{RF}}\mathbf{y}.
		\end{aligned}
	\end{equation}	
	
	Thanks to the diagonal structure of $\mathbf{A}$, (\ref{eq17}) is equivalent to
	\begin{equation}\label{rf}
		\begin{aligned}
			\hat{\mathbf{H}}_\mathrm{RF}=\frac{\mathbf{Y}\odot\mathbf{S}^*}{|\mathbf{S}|^2}=\frac{\mathbf{Y}}{\mathbf{S}},
		\end{aligned}
	\end{equation}
	where $\hat{\mathbf{H}}_\mathrm{RF}$ is the RF estimation of $\mathbf{H}$.
	
	\subsubsection{Matched Filtering (MF)}
	MF is designed based on the rule of maximum SNR output, which may be quantified as 
	\begin{equation}\label{snrout}
		\begin{aligned}
			\mathrm{SNR}_\mathrm{out}= & \frac{\mathbb{E}\{\Vert\mathbf{G}_\mathrm{MF}\mathbf{A}\mathbf{h}\Vert^2_F\}}{\mathbb{E}\{\Vert\mathbf{G}_\mathrm{MF}\mathbf{z}\Vert^2_F\}} = \mathrm{SNR}_\mathrm{in} \cdot \frac{tr\left(\mathbf{G}_\mathrm{MF}\mathbf{A}\mathbf{A}^H\mathbf{G}^H_\mathrm{MF}\right)}{tr\left(\mathbf{G}_\mathrm{MF}\mathbf{G}^H_\mathrm{MF}\right)}
			\\ \leq & \mathrm{SNR}_\mathrm{in}\cdot \frac{tr\left(\mathbf{A}\mathbf{A}^H\right)tr\left(\mathbf{G}^H_\mathrm{MF}\mathbf{G}_\mathrm{MF}\right)}{tr\left(\mathbf{G}_\mathrm{MF}\mathbf{G}^H_\mathrm{MF}\right)}
			\\ = & \mathrm{SNR}_\mathrm{in}\cdot  tr\left(\mathbf{A}\mathbf{A}^H\right)
			= NM \cdot \mathrm{SNR}_\mathrm{in},
		\end{aligned}
	\end{equation}
	where $\mathrm{SNR}_\mathrm{in}=\sum^Q_{q=1}\sigma^2_{\alpha_q}/\sigma^2$, and the maximum $\mathrm{SNR}_\mathrm{out}$ is attained when $\mathbf{G}_\mathrm{MF}=\beta\mathbf{A}^H$. Without loss of generality, we use $\beta=1$, leading to
	\begin{equation}
		\begin{aligned}
			\hat{\mathbf{h}}_\mathrm{MF}=\mathbf{A}^H\mathbf{y}.
		\end{aligned}
	\end{equation} 
	
	Consequently, $\mathbf{H}$ is estimated through
	\begin{equation}\label{mf}
		\begin{aligned}
			\hat{\mathbf{H}}_\mathrm{MF}=\mathbf{Y}\odot\mathbf{S}^*.
		\end{aligned}
	\end{equation}
	
	\subsubsection{Wiener Filtering (WF)}
	WF is designed based on the LMMSE strategy, with its objective to minimize the expectation of squared error as
	\begin{equation}
		\begin{aligned}
			{\mathbf{G}}_\mathrm{WF}=\arg\min_{\mathbf{G}} \ \mathbb{E}\{\Vert\mathbf{h}-\mathbf{G}\mathbf{y}\Vert^2_F\},
		\end{aligned}
	\end{equation}	
	which yields the solution as
	\begin{equation}\label{equ18}
		\begin{aligned}
			\hat{\mathbf{h}}_\mathrm{WF}=\underbrace{\mathbf{R}_h\mathbf{A}^H\left(\mathbf{A}\mathbf{R}_h\mathbf{A}^H+\mathbf{R}_z\right)^{-1}}_{\mathbf{G}_\mathrm{WF}}\mathbf{y}.
		\end{aligned}
	\end{equation}
	
	Therefore, folding (\ref{equ18}) back into matrix form, the LMMSE estimate of imaging channel matrix is
	\begin{equation}\label{wf}
		\begin{aligned}
			\hat{\mathbf{H}}_\mathrm{WF}=\frac{\mathbf{Y}\odot\mathbf{S}^*}{|\mathbf{S}|^2+\mathrm{SNR}^{-1}_\mathrm{in}}.
		\end{aligned}
	\end{equation}
	In practice, the prior information $\mathrm{SNR}_\mathrm{in}$ for WF can be estimated from the received data, e.g., via background noise estimation and signal power measurement. The proposed analysis assumes that such estimates are available.
	
	\vspace{1mm}
	
	Combining (\ref{rf}), (\ref{mf}) and (\ref{wf}), we summarize the TF-domain filtering matrix $\mathbf{G}$ under different schemes as
	\begin{equation}\label{g}
		\begin{aligned}
			\mathbf{G} = 
			\left\{
			\begin{array}{ll}
				\frac{1}{\mathbf{S}} ,   & \mathrm{RF}, \\
				\mathbf{S}^*,   & \mathrm{MF}, \\
				\frac{\mathbf{S}^*}{\left\vert \mathbf{S} \right\vert^2+\mathrm{SNR}_\mathrm{in}^{-1}}, & \mathrm{WF},
			\end{array}
			\right.
		\end{aligned}
	\end{equation}
	and with its $(n,m)$th element expressed as
	\begin{equation}\label{g}
		\begin{aligned}
			g_{n,m} = 
			\left\{
			\begin{array}{ll}
				\frac{1}{s_{n,m}} ,   & \mathrm{RF}, \\
				s_{n,m}^*,   & \mathrm{MF}, \\
				\frac{s_{n,m}^*}{\left\vert s_{n,m} \right\vert^2+\mathrm{SNR}_\mathrm{in}^{-1}}, & \mathrm{WF}.
			\end{array}
			\right.
		\end{aligned}
	\end{equation}
	
	Therefore, the SAR data preprocessed by TF-domain filtering essentially represent an estimated imaging channel matrix, expressed as
	\begin{equation}
		\begin{aligned}
			\mathbf{Y}^\mathrm{tf} \overset{\triangle}{=} \hat{\mathbf{H}} = \mathbf{Y}\odot\mathbf{G} = \mathbf{H}\odot\mathbf{S}\odot\mathbf{G}+\mathbf{Z}\odot\mathbf{G},
		\end{aligned}
	\end{equation}
	where its $(n,m)$th element can be formulated as 
	\begin{equation}\label{equ5}
		\begin{aligned}
			y^\mathrm{tf}_{n,m} = \sum\nolimits_{q}\alpha_q \chi_{n,m} & e^{-j\frac{4\pi}{c}n\Delta f (\Delta R_{q,m}+\bar{R}_q)} \\ & \times e^{-j\frac{4\pi}{c} f_c\Delta R_{q,m}}  +z^\mathrm{tf}_{n,m}.
		\end{aligned}
	\end{equation}
	In (\ref{equ5}), $\chi_{n,m}=s_{n,m}g_{n,m}$ represents the filtered spectrum in the TF domain, and $z^\mathrm{tf}_{n,m}=z_{n,m}g_{n,m}$ is the filtered noise which may be power-amplified due to mismatched filtering.

	\begin{remark}
		The MSE of imaging channel matrix can be derived as
		\begin{equation}\label{equ41}
			\begin{aligned}
				\mathrm{MSE} = & \mathbb{E} 	\left\{\left\Vert\hat{\mathbf{H}}-\mathbf{H}\right\Vert^2_F\right\}
				\\ = & \mathbb{E} \left\{\left\Vert 	\mathbf{H}\odot\mathbf{S}\odot\mathbf{G}-\mathbf{H}+\mathbf{Z}\odot\mathbf{G}\right\Vert^2_F\right\}
				\\ = & NM\left(\sum\nolimits_{q} \sigma^2_{\alpha_q} \mathbb{E}\{(\chi-1)^2 \}  +\sigma^2\mathbb{E}\{|g|^2\} \right),
			\end{aligned}
		\end{equation}
		where the third equality holds due to the facts that: 1) the statistical independence between $\mathbf{H}$ and $\mathbf{Z}$, and 2) the statistical characteristics of random variables in (\ref{equ41}) are invariant across subcarriers and symbols. Readers are referred to  \cite{du2025prob} for a detailed derivation.
	\end{remark}

	\subsection{Range Compression}
	Evidently, the range (delay)-induced phase is linear within the frequency domain, which is different from standard radar waveforms such as chirp. Therefore, the range compression can be straightforwardly achieved via an IFFT operation accordingly, leading to SAR data mapped in the range-azimuth domain as
	\begin{equation}\label{equ6}
		\begin{aligned}
			y^\mathrm{rc}_{k,m} = & \frac{1}{\sqrt{N}}\sum\nolimits_{n} y^\mathrm{tf}_{n,m} e^{j2\pi \frac{nk}{N}} 
			\\ = & \sum\nolimits_{q}\alpha_q \underbrace{\left[\frac{1}{\sqrt{N}}\sum\nolimits_{n}\chi_{n,m}e^{j\frac{2\pi n}{N}\left(k-k_{q,m}\right)}\right]}_{r_m\left(k-k_{q,m}\right)} \\ & \qquad \times e^{-j\frac{4\pi}{c} f_c\Delta R_{q,m}} +z^\mathrm{rc}_{k,m},
		\end{aligned}
	\end{equation}
	where $\rho_r=\frac{c}{2B_r}$ is the range resolution, and
	\begin{equation}\label{equ35}
		\begin{aligned}
			k=k_{q,m}&=(\Delta R_{q,m}+\bar{R}_q)/\rho_r \\ & = \left(\frac{(vT_{\mathrm{sym}}m-y_q)^2}{2\bar{R}_q}+\bar{R}_q\right)/\rho_r.
		\end{aligned}
	\end{equation}
	
	Moreover, the shapes of RCM corresponding to different azimuths but identical ranges are the same, while they may differ with same azimuths at separate ranges. 
	In fact, all targets' trajectories can be approximately aligned, which will be demonstrated subsequently. This motivates the azimuth FFT operation below.

	\subsection{{Azimuth FFT}}	
	Let us perform FFT in the azimuth domain, which yields the SAR data in the RD domain as
	\begin{equation}\label{equ7}
		\begin{aligned}
			y^\mathrm{rd}_{k,p} = & \frac{1}{\sqrt{M}} \sum\nolimits_{m} y^\mathrm{rc}_{k,m} e^{-j2\pi \frac{mp}{M}} 
			\\ = & \frac{1}{\sqrt{M}} \sum\nolimits_{q}\alpha_q  \sum\nolimits_{m} {r_m\left(k-k_{q,m}\right)} e^{-j\frac{4\pi}{c} f_c\Delta R_{q,m}} \\ & \times e^{-j2\pi \frac{mp}{M}} +z^\mathrm{rd}_{k,p}.
		\end{aligned}
	\end{equation}
	
	In the following, the azimuth phase term in (\ref{equ7}) is decomposed into multiple parts for further analysis, expressed as
	\begin{equation}\label{equ36}
		\begin{aligned}
			e^{-j\frac{4\pi}{c} f_c\Delta R_{q,m}} = & e^{-j2\pi\frac{ v^2T^2_\mathrm{sym}}{\lambda\bar{R}_q}m^2} e^{j4\pi\frac{vT_{\mathrm{sym}}y_q}{\lambda \bar{R}_q}m}  e^{-j2\pi\frac{ y^2_q}{\lambda\bar{R}_q}}
			\\ = & e^{-j\pi K_a T^2_{\mathrm{sym}}m^2} e^{j2\pi \frac{K_aT_{\mathrm{sym}}y_q}{v}m} e^{-j2\pi\frac{ y^2_q}{\lambda\bar{R}_q}},
		\end{aligned}
	\end{equation}
	where $K_a=\frac{2v^2}{\lambda\bar{R}_q}$ denote the slope frequency in the azimuth direction. Clearly, a quadratic azimuth phase appears, exhibiting as an azimuth chirp. Notably, the amplitude envelope $r_m\left(k-k_{q,m}\right)$ in (\ref{equ7}) varies over time, arising mainly from two reasons: 1) the target’s RCM induces amplitude modulation in the azimuth signal, and 2) the spectrum shaping resulting from TF-domain filtering (i.e., $\chi_{n,m}$) further affects the temporal envelope. Therefore, it is not straightforward to derive the azimuth spectrum in (\ref{equ7}). 
	
	To that end, we may exploit the celebrated principle of stationary phase approximation (SPA) \cite{cumming2005digital}, to approach the azimuth Doppler spectrum as demonstrated in (\ref{equ7}). However, the existing studies mainly focus on spectral approximations for continuous chirp signals, with no reports on the approximations of discrete OFDM communication signals modulated by non-constant modulus constellation symbols. For the completeness of this article, we develop its discrete form associated with the premise of SPA, as summarized below.
	
	\begin{lemma}
		The azimuth envelop $r_m\left(k-k_{q,m}\right)$ varies much slower  than the phase of $e^{-j\frac{4\pi}{c} f_c\Delta R_{q,m}}$.
	\end{lemma}
	
	\textbf{\textit{Proof:}} 
	See Appendix \ref{appendix1}.
	\hfill $\blacksquare$
	\vspace{3mm}

	\begin{theorem}
		For the discrete signal $u_m=w_me^{j\phi(m)}$, when the envelope $w_m$ varies much slower than the quadratic‐phase (chirp‐like) $\phi(m)=am^2+bm+\mathrm{constant}$, the FFT of $u_m$ can be approximated as
		\begin{equation}\label{theorem1}
			\begin{aligned}
				U_p = & \sum\nolimits_m w_m e^{j\phi(m)}e^{-j2\pi mp/M}
				\\ \approx & |w_{\tilde{m}}| e^{j\Phi(\tilde{m})} \sqrt{\frac{2\pi}{\left\vert\Phi''(\tilde{m})\right\vert}}e^{j \mathrm{sgn}\left(\Phi''(\tilde{m})\right)\frac{\pi}{4}},
			\end{aligned}
		\end{equation}
		where $\Phi(m)=\phi(m)-2\pi mp/M + \arg w_m$, and $\tilde{m}$ denotes the stationary point obtained according to $\frac{d}{dm}\Phi(m)|_{\tilde{m}}=0$, expressed as
		\begin{equation}\label{posp}
			\begin{aligned}
				\tilde{m}\approx \frac{1}{2a}\left(\frac{2\pi p}{M}-b\right).
			\end{aligned}
		\end{equation}
	\end{theorem}
	
	\textbf{\textit{Proof:}} 
	See Appendix \ref{appendix2}.
	\hfill $\blacksquare$
	\vspace{3mm}


	Combining (\ref{posp}) and (\ref{equ36}) with $a=-\pi K_a T^2_{\mathrm{sym}}$ and $b=2\pi \frac{K_ay_qT_{\mathrm{sym}}}{v}$, the stationary point is derived as
	\begin{equation}\label{posp2}
		\begin{aligned}
			\tilde{m}\approx -\frac{p}{MT^2_{\mathrm{sym}}K_a}+\underbrace{\frac{y_q}{vT_{\mathrm{sym}}}}_{m_q} \in [0,M)
		\end{aligned}
	\end{equation}
	where the range of index $p$ can thus be determined accordingly.
	
	Combining (\ref{posp2}), (\ref{theorem1}) and (\ref{equ7}), and performing a series of algebraic manipulations and simplifications, the azimuth Doppler spectrum can be approximated as
	\begin{equation}\label{equ8}
		\begin{aligned}
			& \sum\nolimits_{m}{r_m\left(k-k_{q,m}\right)} e^{-j\frac{4\pi}{c} f_c\Delta R_{q,m}} e^{-j2\pi \frac{mp}{M}} \\
			& \approx \sqrt{M} \varepsilon r_{\tilde{m}}\left(k-k_{q,\tilde{m}}\right)e^{j\pi\frac{p^2}{M^2T^2_{\mathrm{sym}}K_a}} e^{-j2\pi\frac{m_qp}{M}}  e^{-j2\pi\frac{ y^2_q}{\lambda\bar{R}_q}},
		\end{aligned}
	\end{equation}
	where $e^{j \mathrm{sgn}\left(\Phi''(\tilde{m})\right)\frac{\pi}{4}}$ is omitted in (\ref{equ8}) as it does not affect the SAR imaging quality. Notice that the coefficient $\sqrt{M}\varepsilon$ in (\ref{equ8}) is induced from the following equation:
	\begin{equation}
		\begin{aligned}
			\sqrt{\frac{2\pi}{\left\vert\Phi''(\tilde{m})\right\vert}}=\sqrt{\frac{2\pi}{|2a|}}=\sqrt{\frac{1}{K_a T^2_{\mathrm{sym}}}}=\sqrt{M} \varepsilon.
		\end{aligned}
	\end{equation}
	Here, we define the last equality sign in order to cancels the coefficient $1/\sqrt{M}$ in (\ref{equ7}), and
	\begin{equation}
		\begin{aligned}
			\varepsilon=\sqrt{\frac{1}{K_a M T^2_{\mathrm{sym}}}}=\sqrt{\frac{\mathrm{PRF}}{B_a}}\geq 1,
		\end{aligned}
	\end{equation} 
	where $\mathrm{PRF}=1/T_\mathrm{sym}$ is the azimuth sampling rate (i.e., pulse repetition frequency) and $B_a=K_aMT_\mathrm{sym}$ is the azimuth bandwidth.
	Interestingly, using the SPA introduces an artificial amplitude ``gain'' $\varepsilon$ in the approximated frequency-domain result. However, in reality, the true signal power (and thus SNR) is not increased, as this apparent ``gain'' is purely an artifact of the approximation, which is unrealistic, since the FFT is a unitary transform that does not change the total power of the signal or noise. Due to this reason, below we let $\varepsilon=1$ to avoid misleading.
	
	Finally, by plugging (\ref{equ8}) into (\ref{equ7}), we arrive at
	\begin{equation}\label{equ9}
		\begin{aligned}
			y^\mathrm{rd}_{k,p} \approx \sum\nolimits_{q} \tilde{\alpha}_q r_{\tilde{m}}\left(k-k_{q,\tilde{m}}\right)  e^{j\pi\frac{p^2}{M^2T^2_{\mathrm{sym}}K_a}} e^{-j2\pi\frac{m_qp}{M}}  + z^\mathrm{rd}_{k,p},
		\end{aligned}
	\end{equation}
	where $\tilde{\alpha}_q=\alpha_qe^{-j2\pi\frac{ y^2_q}{\lambda\bar{R}_q}}$.
	
	\begin{remark}
		The migration trajectories in the RD domain can be expressed as
		\begin{equation}\label{equ38}
			\begin{aligned}
				k & =k_{q,\tilde{m}}= (\Delta 	R_{q,\tilde{m}}+\bar{R}_q)/\rho_r \\ & =  \left(\frac{(vT_{\mathrm{sym}}\tilde{m}-y_q)^2}{2\bar{R}_q}+\bar{R}_q\right)/\rho_r
				\\ & =  	\left(\frac{v^2}{2\bar{R}_q(K_aMT_{\mathrm{sym}})^2}p^2+\bar{R}_q\right)/\rho_r,
			\end{aligned}
		\end{equation}
		where the last equality sign holds by substituting (\ref{posp2}) into (\ref{equ38}). This result demonstrates that: 1) RCM has no relationship with $y_q$, and 2) the quadratic coefficient in (\ref{equ38}), i.e., $\frac{v^2}{2\bar{R}_q(K_aMT_{\mathrm{sym}})^2}$, is much less affected by $\bar{R}_q$, relative to $\frac{v^2T^2_{\mathrm{sym}}}{2\bar{R}_q}$ in (\ref{equ35}). These findings suggest that shapes of all RCM trajectories are almost same in the RD domain.
 		Consequently, we may conduct RCM correction (RCMC) by applying a Doppler-dependent range resampling (interpolation) scale, so that the target energy, which has migrated across range bins as a function of Doppler, can be realigned onto constant range cells.
		This operation significantly reduces the complexity relative to grid-wise compensation in the range-azimuth domain.
	\end{remark}

	\subsection{{RCMC}}
	In order to compensate for the RCM, for each Doppler frequency, we estimate the target range offset relative to a reference range, and then resample the range profile to shift it by using a complex interpolation kernel. Below we detail the RCMC with sinc interpolation, where the interpolated SAR data is given by
	\begin{equation}
		\begin{aligned}
			y^\mathrm{rcmc}_{k,p} = y^\mathrm{rd}_{k+\Delta k,p} 
			\approx & \sum\nolimits_{k'} y^\mathrm{rd}_{k',p}\mathrm{sinc}\left(k'-\left(k+\Delta k\right)\right),
		\end{aligned}
	\end{equation}
	where $\Delta k$ denotes the RCM which needs to be compensated for each Doppler column, and can be computed as $\Delta k = \Delta R_{q,\tilde{m}}/\rho_r = \frac{v^2}{2\bar{R}_q(K_aMT_{\mathrm{sym}})^2\rho_r}p^2$, by referring to (\ref{equ38}). 
	
	If we assume perfect interpolations\footnote{In practice, truncating the sinc kernel limits its accuracy. It may cause slight resolution loss, sidelobe distortion, and residual RCM errors \cite{mao2018knowledge}, especially for large fractional shifts. As we focus on the impact of signaling randomness on SAR imaging, RCM is thus assumed to be perfectly corrected.}, then Doppler columns contain target energy aligned to the same range index, giving rise to
	\begin{equation}\label{equ10}
		\begin{aligned}
			y^\mathrm{rcmc}_{k,p} \approx \sum\nolimits_{q} \tilde{\alpha}_q r_{\tilde{m}}\left(k-k_{q}\right)   e^{j\pi\frac{p^2}{M^2T^2_{\mathrm{sym}}K_a}} e^{-j2\pi\frac{m_qp}{M}}  + z^\mathrm{rcmc}_{k,p},
		\end{aligned}
	\end{equation}
	where $r_{\tilde{m}}\left(k-k_{q}\right)$ can be reconstructed as
	\begin{equation}
		\begin{aligned} 
			r_{\tilde{m}}\left(k-k_q\right) = \frac{1}{\sqrt{N}}\sum\nolimits^{N-1}_{n=0}\chi_{n,\tilde{m}}e^{j\frac{2\pi n}{N}\left(k-k_q\right)},
		\end{aligned}
	\end{equation}
	and  $k_q=\bar{R}_q/\rho_r$, which demonstrates that the RCM in (\ref{equ38}) has been entirely compensated.

	\begin{remark}
		The noise $z^\mathrm{rcmc}_{k,p}$ also experiences the sinc interpolation, expressed as
		\begin{equation}
			\begin{aligned}
				z^\mathrm{rcmc}_{k,p} = 
				\sum\nolimits_{k'} z^\mathrm{rd}_{k',p}\mathrm{sinc}\left(k'-\left(k+\Delta k\right)\right).
			\end{aligned}
		\end{equation}
		Then we may verify the statistic characteristic of $z^\mathrm{rcmc}_{k,p}$ as
			\begin{align}
				& \mathbb{E}\{z^\mathrm{rcmc}_{k,p}\} =  0
				\\ 
				&\mathbb{E}\left\{z^\mathrm{rcmc}_{k,p}\left(z^\mathrm{rcmc}_{\tilde{k},p}\right)^*\right\} \\ & = \sigma^2\mathbb{E}\{|g|^2\}
				\sum_{k'} \mathrm{sinc}\left(k'-\left(k+\Delta k\right)\right)  \mathrm{sinc}\left[k'-\left(\tilde{k}+\Delta k\right)\right].\notag
			\end{align}
		Notably, sinc interpolation preserves the Gaussian nature of noise but makes it slightly correlated, since each output is a weighted sum of multiple noisy inputs when $\Delta k$ is not an integer. However, this minor effect is typically negligible in practice.
	\end{remark}

	\subsection{{Azimuth Compression}}
	Since the trajectories of all targets are straightened after RCMC, it is straightforward to see that compensating for the Doppler chirp phase and performing IFFT can achieve azimuth compression, leading to 
	\begin{equation}\label{equ11}
		\begin{aligned}
			& y^\mathrm{ac}_{k,m} = \frac{1}{\sqrt{M}}\sum\nolimits_{p} y^\mathrm{rcmc}_{k,p} e^{-j\pi\frac{p^2}{M^2T^2_{\mathrm{sym}}K_a}} e^{j2\pi\frac{mp}{M}}
			\\ & = \frac{1}{\sqrt{M}} \sum\nolimits_{q}\tilde{\alpha}_q \sum\nolimits_{p} r_{\tilde{m}} \left(k-k_{q}\right)  e^{j\pi\frac{p^2}{M^2T^2_{\mathrm{sym}}K_a}}e^{-j\pi\frac{p^2}{M^2T^2_{\mathrm{sym}}K_a}} \\ & \quad \times e^{-j2\pi\frac{m_qp}{M}} e^{j2\pi\frac{mp}{M}} + z^\mathrm{ac}_{k,m} 
			\\ & = \sum_{q}\tilde{\alpha}_q \underbrace{\left[\frac{1}{\sqrt{NM}} \sum_{n,p} \chi_{n,\tilde{m}} e^{j2\pi\frac{n\left(k-k_q\right)}{N}}e^{j2\pi\frac{p\left(m-m_q\right)}{M}}\right] }_{R(k-k_q,m-m_q)}+ z^\mathrm{ac}_{k,m},
		\end{aligned}
	\end{equation}
	where $R(k,m)$ represents the point spread function (PSF), which can be interpreted as a target response function in the range-azimuth domain.
	
	\vspace{2mm}
	Overall, the proposed OFDM-SAR imaging approach remains consistent with the conventional RD algorithm, which is applicable for chirp signaling.
	However, prior to the range compression, a random symbol-dependent TF-domain filtering scheme is preprocessed to mitigate the influence of embedded data symbols from echoes. As such, the imaging quality may be influenced by both the digital modulation schemes and TF-domain filtering strategies, unlike conventional SAR imaging. Therefore, it is necessary to quantitatively analyze the OFDM-SAR imaging quality and introduce new performance evaluation metrics, in particular to illustrate the impact of signaling randomness.


	\section{Mapping Mechanism Between Data Randomness and SAR Imaging}\label{sec4}	
	\subsection{Conventional SAR Imaging Metrics}
	Practical SAR measurements typically deploy a corner reflector within the imaging scene and use its reconstructed response as the reference PSF \cite{chatzitheodoridi2022cooperative}. Based on the PSF, a variety of point-target performance metrics, such as range-azimuth resolution, integrated sidelobe ratio (ISLR) \cite{stoica2009new}, noise-equivalent sigma zero (NESZ) \cite{patyuchenko2009design}, and peak energy loss (PEL) \cite{du2025prob}, have been introduced. By examining the imaging quality of this corner reflector, these metrics provide a standardized and comparable basis across different SAR systems and processing algorithms. In this context, the evaluation metrics of PSF generally fall into three main categories:
	
	\subsubsection{Resolution‐related (mainlobe) indicators}
	Range and azimuth resolutions are two key metrics to distinguish two closest points in a reconstructed image. Specifically, the range resolution $\rho_r$ depends on the signaling bandwidth $B_r$, i.e.,
	\begin{equation}\label{eq48}
		\begin{aligned}
			\rho_r = {c}/(2B_r),
		\end{aligned}
	\end{equation}
	which has been mentioned previously.
	In contrast, the azimuth resolution $\rho_a$ depends on the synthetic aperture bandwidth $B_a=K_aT_a$, expressed as
	\begin{equation}\label{eq49}
		\begin{aligned}
			\rho_a = v/{B_a}.
		\end{aligned}
	\end{equation}
	
	Observing (\ref{eq48}) and (\ref{eq49}), it is naturally to see that both resolutions are irrelevant to the signaling randomness. However, they may differ between pilot-only imaging and data-aided imaging schemes, since pilots are sparsely inserted in the frame structure with a much smaller resource occupation. This would be validated in Sec.~\ref{sec5}.
	
	\subsubsection{Sidelobe indicators} The ISLR is a popular imaging metric used to quantify the total energy of the sidelobes relative to the mainlobe of PSF \cite{stoica2009new,liu2017circulate,zhen2021multicarrier}. In the context of the signaling randomness in OFDM-SAR imaging, ISLR can be defined as 
	\begin{equation}\label{eq31}
		\begin{aligned}
			\mathrm{ISLR} = \frac{\mathbb{E} \left\{\sum\nolimits_{k,m} |R(k,m)|^2\right\} - \mathbb{E} \left\{R^2(0,0)\right\}}{\mathbb{E} \left\{R^2(0,0)\right\}}.
		\end{aligned}
	\end{equation}
	
	Overall, sidelobe indicators are primarily designed to characterize the PSF shape under well-focused conditions, and thus do not explicitly account for noise effects. In conventional SAR systems, noise-related performance is typically assessed using separate radiometric metrics such as NESZ or output SNR. Therefore, rather than indicating a limitation of ISLR, this reflects its specific role in evaluating sidelobe behavior.

	\subsubsection{Radiometric quality indicators} According to \cite{du2025prob}, mismatched filtering schemes may arise an additional PEL of PSF, and amplify the noise power. To illustrate this, we may resort to the following metrics.
		
		$\bullet$ \textbf{PEL}: In practical SAR imaging systems, factors such as windowing, interpolation errors, truncation of the PSF kernel, motion errors, or waveform distortions can reduce the peak energy of PSF. The PEL thus serves as a measure of the system’s ability to preserve target amplitude and contrast in the reconstructed image. As we concentrate on the impact of signaling randomness on the imaging performance, we notice that both MF and RF preserve the peak energy of PSF while WF induces a peak loss, due to its scale-variant characteristic. By referring to \cite{du2025prob}, we may define PEL as
		\begin{equation}
			\begin{aligned}
				\mathrm{PEL} = NM\mathbb{E} \left\{\left(1-{R(0,0)}/{\sqrt{NM}}\right)^2\right\},
			\end{aligned}
		\end{equation}
		where $NM$ is the range-azimuth compression gain.
		
		$\bullet$ \textbf{Output SNR}: In SAR systems, the NESZ \cite{patyuchenko2009design} represents the RCS of a reference target when it equals the system noise. In other words, NESZ reflects the minimum detectable scattering, i.e., the system sensitivity, when $\mathrm{SNR_{out}= 0 \ dB}$. Note that $\text{SNR}_\text{out}$ directly represents the ratio of the mainlobe peak energy of PSF to the output noise power. Therefore, it inherently incorporates NESZ, and further accounts for the target’s scattering strength and focusing gain, thus making it a more intuitive measure of the visibility and imaging quality. $\text{SNR}_\text{out}$ has been defined in (\ref{snrout}), which is equivalent to the following expression in terms of the PSF \cite{du2025prob}:
		\begin{equation}\label{snr}
			\begin{aligned}
				\mathrm{SNR_{out}}  = & \frac{\sigma^2_{\alpha} \mathbb{E}\left\{R^2(0,0)  \right\}}{\sigma^2 \mathbb{E}\left\{|g|^2 \right\} }.
			\end{aligned}
		\end{equation}

	\subsection{OFDM-SAR Imaging Metrics}
	In scenarios where both sidelobe structure and noise are affected by the modulation randomness and TF-filtering schemes, it is desirable to adopt a unified metric that jointly captures these effects. To this end,
	one may exploit the MSE of the entire SAR profile as a possible imaging metric. However, it is generally not practical, since the ground-truth RCS distribution of all scatterers is unknown in real scenes, making it impossible to compute the grid/pixel-wise error between the reconstructed image and the ideal reference. Moreover, the large dynamic range of RCS values causes the MSE to be dominated by a few strong scatterers, which fails to reflect the regional image quality\footnote{To address this limitation, the structural similarity index measure (SSIM) \cite{wang2004image} is used to evaluate image quality in a perceptually meaningful way. Instead of measuring absolute intensity deviations, SSIM compares local patterns of luminance, contrast, and structure between two images within a sliding window. This normalization suppresses the dominance of strong reflections and allows for a more balanced assessment of structural fidelity across different intensity levels. Nevertheless, both MSE and SSIM are reference-dependent metrics that require a ground-truth image for comparison.}.
	Therefore, SAR image quality is typically assessed using physical or statistical indicators that can be directly derived from the reconstructed image, such as the above mentioned point target-related metrics, which are free of reference images.
	 
	Nevertheless, in this subsection, we demonstrate that the normalized MSE (NMSE) of a known reference point target (i.e., the corner reflector) can serve as a relatively comprehensive and practically relevant SAR imaging metric, and elucidate its relationship with the conventional metrics. To proceed, we first derive the average imaging profile, i.e., the expectation of the squared modulus of (\ref{equ11}), expressed as 
	\begin{equation}\label{equ12}
		\begin{aligned}
			\mathbb{E}  & \{|y^\mathrm{ac}_{k,m}|^2\} = \sum_q \sigma^2_{\alpha_q}\frac{1}{NM}\sum\nolimits_{n,p}\sum\nolimits_{n',p'}\mathbb{E}\{\chi_{n,\tilde{m}}\chi_{n',\tilde{m}'} \} \\ & \times e^{j2\pi \frac{(n-n')(k-k_q)}{N}}e^{j2\pi \frac{(p-p')(m-m_q)}{M}} + \mathbb{E}\{z^\mathrm{ac}_{k,m}\left(z^\mathrm{ac}_{k,m}\right)^*\}, 
		\end{aligned}
	\end{equation}
	where $\mathbb{E}\{\tilde{\alpha}_q\tilde{\alpha}_{q'}^* \}_{q\neq q'}=0$ and $\mathbb{E}\{\tilde{\alpha}_q (z^\mathrm{ac}_{k,m})^*\}=0$ are utilized for simplicity.
	
	For a further simplification, the first term on the right side of (\ref{equ12}) can be derived as (\ref{equ13}) at the top of this page. During this derivation, Dirichlet kernels are exploited \cite{du2025prob}, expressed as $\sum_{n,n'} e^{j\frac{2\pi(n-n')(k-k_q)}{N}} \approx N^2\mathrm{sinc}^2(k-k_q)$ and $\sum_{p,p'} e^{j\frac{2\pi(p-p')(m-m_q)}{M}}  \approx M^2\mathrm{sinc}^2(m-m_q)$. 
	\begin{figure*}[!t]
		\begin{equation}\label{equ13}
			\begin{aligned}
				& \sum\nolimits_{n,p}\sum\nolimits_{n',p'}\mathbb{E}\{\chi_{n,\tilde{m}}\chi_{n',\tilde{m}'} \} e^{j2\pi \frac{(n-n')(k-k_q)}{N}}e^{j2\pi \frac{(p-p')(m-m_q)}{M}} 
				\\ & = \sum\nolimits_{n,p}\sum\nolimits_{n',p'}\mathbb{E}\{\chi_{n,\tilde{m}}\}\mathbb{E}\{\chi_{n',\tilde{m}'}\} e^{j2\pi \frac{(n-n')(k-k_q)}{N}} e^{j2\pi \frac{(p-p')(m-m_q)}{M}} 
				+ \sum\nolimits_{n,p}\mathbb{E}\{\chi_{n,\tilde{m}}^2 \} - \sum\nolimits_{n,p}\mathbb{E}^2\{\chi_{n,\tilde{m}}\} 
				\\ &  = N^2M^2\mathbb{E}^2\{\chi\}\mathrm{sinc}^2\left(k-k_q\right)\mathrm{sinc}^2\left(m-m_q\right)+ NM\underbrace{\left(\mathbb{E}\{\chi^2 \} - \mathbb{E}^2\{\chi\}\right)}_{{\mathrm{Var}}\left(\chi\right)} 
			\end{aligned}
		\end{equation}
		\rule{18cm}{1.0pt}
	\end{figure*}	
	In addition, it is natural to see 
	\begin{equation}\label{equ150}
		\begin{aligned}
			\mathbb{E}\{z^\mathrm{ac}_{k,m}\left(z^\mathrm{ac}_{k,m}\right)^*\}= \sigma^2\mathbb{E}\{|g|^2\}.  
		\end{aligned}
	\end{equation}

	Finally, substituting (\ref{equ13}) and (\ref{equ150}) into (\ref{equ12}), one may obtain
	\begin{equation}\label{eq14}
		\begin{aligned}
			\mathbb{E}& \{|y^\mathrm{ac}_{k,m}|^2\} =  NM\sum\nolimits_q\sigma^2_{\alpha_q}\mathbb{E}^2\{\chi\}\mathrm{sinc}^2\left(k-k_q\right) \\ & \times \mathrm{sinc}^2\left(m-m_q\right) + \sum\nolimits_q\sigma^2_{\alpha_q}\mathrm{Var}\left(\chi\right) + \sigma^2\mathbb{E}\{|g|^2\}.
		\end{aligned}
	\end{equation}

		Evidently, the average SAR imaging profile in (\ref{eq14}) demonstrates the imaging quality is determined by three factors: $\mathbb{E}^2\{\chi\}$, $\mathrm{Var}\left(\chi\right)$ and $\mathbb{E}\{|g|^2\}$. 
		First, $\sum_q\sigma^2_{\alpha_q}\mathrm{Var}\left(\chi\right) $ and $\sigma^2\mathbb{E}\{|g|^2\}$ constitute the constant pedestal in the range-azimuth profile, which are attributed to the signaling randomness and the amplified noise. Second, the expectation of PSF expressed as $\mathbb{E}\{\chi\}\text{sinc}\left(k-k_q\right)\text{sinc}\left(m-m_q\right)$, may be smaller than the ideal one $\text{sinc}\left(k-k_q\right)\text{sinc}\left(m-m_q\right)$ due to $\mathbb{E}\{\chi\}\leq 1$ for MF, RF, and WF. Consequently, it is essential to develop a unified metric to quantify these effects and integrate it with the aforementioned conventional indicators.
		
		\vspace{2mm}
		In addition, an ideal SAR reference image should be independent of both signal randomness and noise, expressed as
		\begin{equation}\label{eq15}
			\begin{aligned}
				y^\mathrm{ideal}_{k,m} = & \frac{1}{\sqrt{NM}} \sum\nolimits_q\tilde{\alpha}_q \sum\nolimits_{n,p} e^{j2\pi\frac{n\left(k-k_q\right)}{N}}e^{j2\pi\frac{p(m-m_q)}{M}}
				\\ = & \sqrt{NM} \sum\nolimits_q\tilde{\alpha}_q \mathrm{sinc}\left(k-k_q\right)\mathrm{sinc}\left(m-m_q\right).
			\end{aligned}
		\end{equation}
		Then the MSE between $y^\mathrm{ac}_{k,m}$ and $y^\mathrm{ideal}_{k,m}$ can be derived as 
		\begin{equation}\label{equ40}
			\begin{aligned}
				\mathrm{MSE} & = \sum\nolimits_{k,m}\mathbb{E}\left\{\left\vert y^\mathrm{ac}_{k,m}-y^\mathrm{ideal}_{k,m} \right\vert^2\right\}
				\\ & = NM\left(\sum\nolimits_q \sigma^2_{\alpha_q} \mathbb{E}\{(\chi-1)^2 \}  +\sigma^2\mathbb{E}\{|g|^2\} \right)
				\\ & = (\ref{equ41}),
			\end{aligned}
		\end{equation}
		where the mathematical derivation and simplification are similar to those of (\ref{eq14}). In addition, the result in (\ref{equ40}) implies that the MSE of SAR profiles is equivalent to the MSE of imaging channel matrix. This is straightforward, since the range and azimuth compression operations are linear and distortion-free, on the premise of a perfect RCMC.

	

	\begin{remark}
		Despite the fact that the MSE of an overall SAR imaging profile is not practical, its normalized counterpart corresponding to a point target (i.e., the corner reflector) can still serve as a meaningful indicator, as the RCS power $\sigma^2_{\alpha}$ of the corner reflector is known in advance. In this case, the relationship among the NMSE, ISLR, PEL, and $\text{SNR}_\text{out}$ can be readily established as
		\begin{equation}\label{theorem}
			\begin{aligned}
				\mathrm{ISLR}+ \frac{\mathrm{PEL}}{\mathbb{E} \left\{R^2(0,0)\right\}} + \frac{1}{\mathrm{SNR_{out}}}  \overset{Q=1}{=} \underbrace{\frac{\mathrm{MSE}}{\sigma^2_{\alpha} \mathbb{E} \left\{R^2(0,0)\right\}}}_{\mathrm{NMSE}},
			\end{aligned}
		\end{equation}
		for which a detailed proof can be found in \cite[Theorem 1]{du2025prob}. In the considered OFDM-SAR framework, NMSE provides a unified measure that jointly reflects the impact of sidelobe structure, peak distortion, and noise, which are otherwise characterized separately by ISLR, PEL, and SNR-related metrics.
		
		It is worth noting that NMSE is not intended to replace conventional SAR performance metrics such as resolution, ISLR, or radiometric indicators. Instead, it serves as a complementary metric that is particularly useful in scenarios where a reference point target is available, such as calibration experiments or controlled simulations. 
		For example, to realize trade-off between sensing and communications via waveform design, NMSE can be directly optimized through probabilistic constellation shaping (PCS) \cite{du2024reshaping}, offering a much simpler alternative to optimizing multiple conventional metrics such as ISLR, PEL, and $\mathrm{SNR_{out}}$. We refer to \cite{du2025prob} for more details.
		Therefore, it is suitable for comparative performance evaluation of different waveform and filtering strategies, and should be interpreted together with conventional SAR metrics for a comprehensive assessment of imaging quality. 
	\end{remark}


	\section{Simulations and Performance Analysis}\label{sec5}
	We consider a UAV-SAR system in LAWNs, where the OFDM signaling parameters follow standard 5G NR configuration at the n78 frequency band, where the carrier frequency is ${f}_c=3.5$ GHz, the bandwidth is $B_r=100$ MHz, the subcarrier spacing is $\Delta f=30$ KHz, the cyclic prefix duration is ${T_{\rm{cp}}}=  8.33 \ \mu s$, the transmit OFDM symbol duration is ${T_p} =33.33 \ \mu s$, and the synthetic azimuth time $T_a =2$ s. Throughout simulations, all data symbols are i.i.d. drawn from a $256$-QAM constellation. Besides, the UAV-SAR platform, which moves at the constant speed of $v=50$ m/s along the azimuth direction, is at the height of $H_p=1000$ m. Notice that the data backhaul follows the standard communication transceiver chain, and thus we merely concentrate on the SAR imaging performance.

	\begin{figure*}[!t]
		\centering 
		\subfigure[Range compression]{
			\includegraphics[width=1.65in]{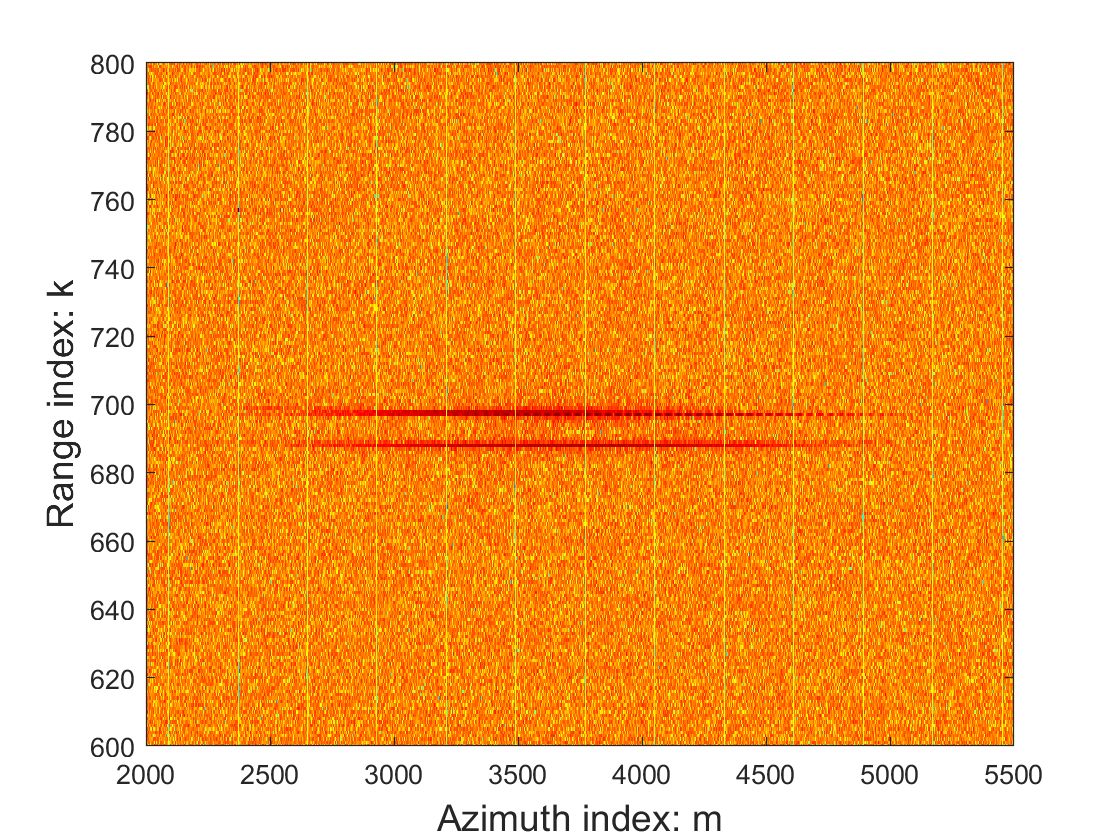}}
		\subfigure[Azimuth FFT]{
			\includegraphics[width=1.65in]{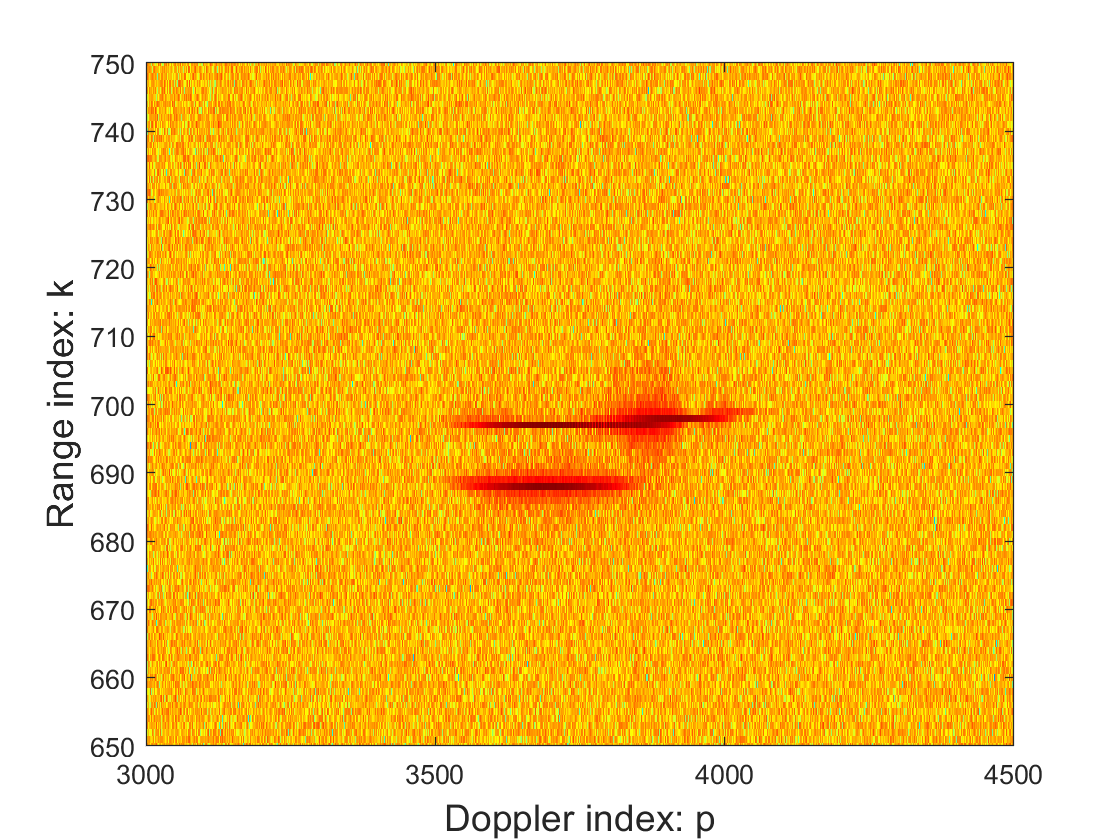}}
		\subfigure[RCMC]{
			\includegraphics[width=1.65in]{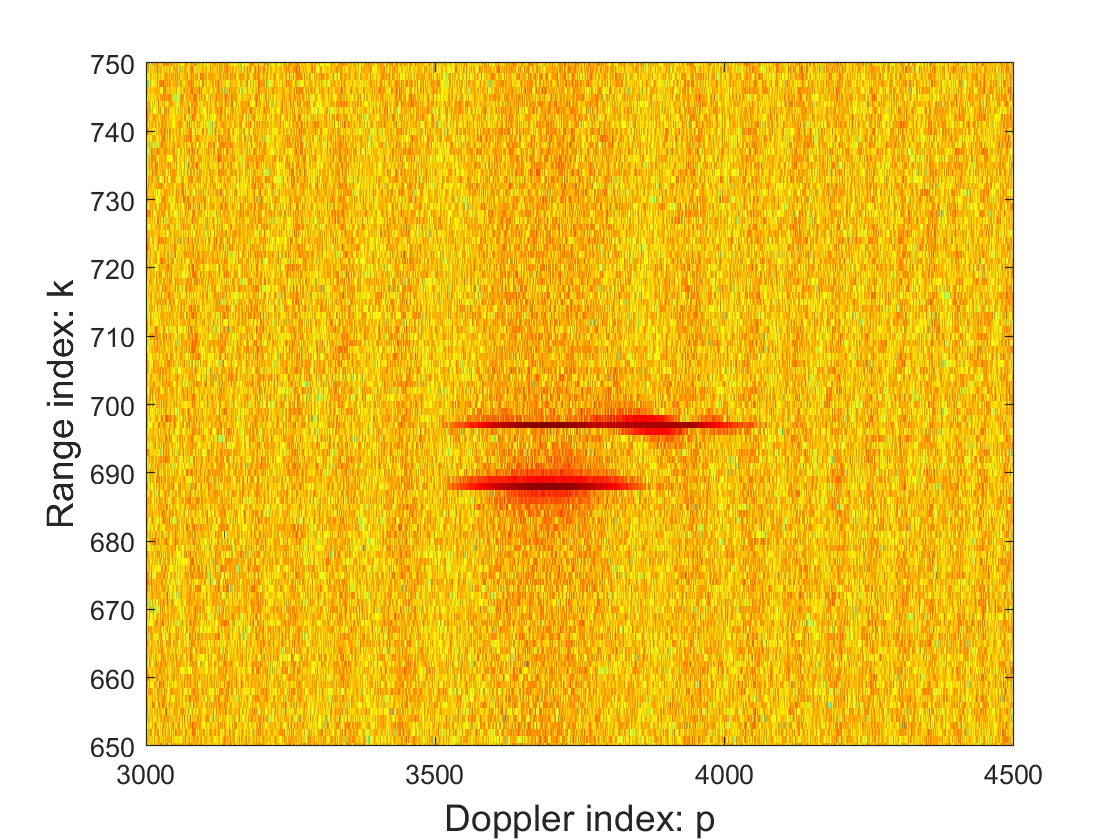}}	
		\subfigure[Azimuth compression]{
			\includegraphics[width=1.65in]{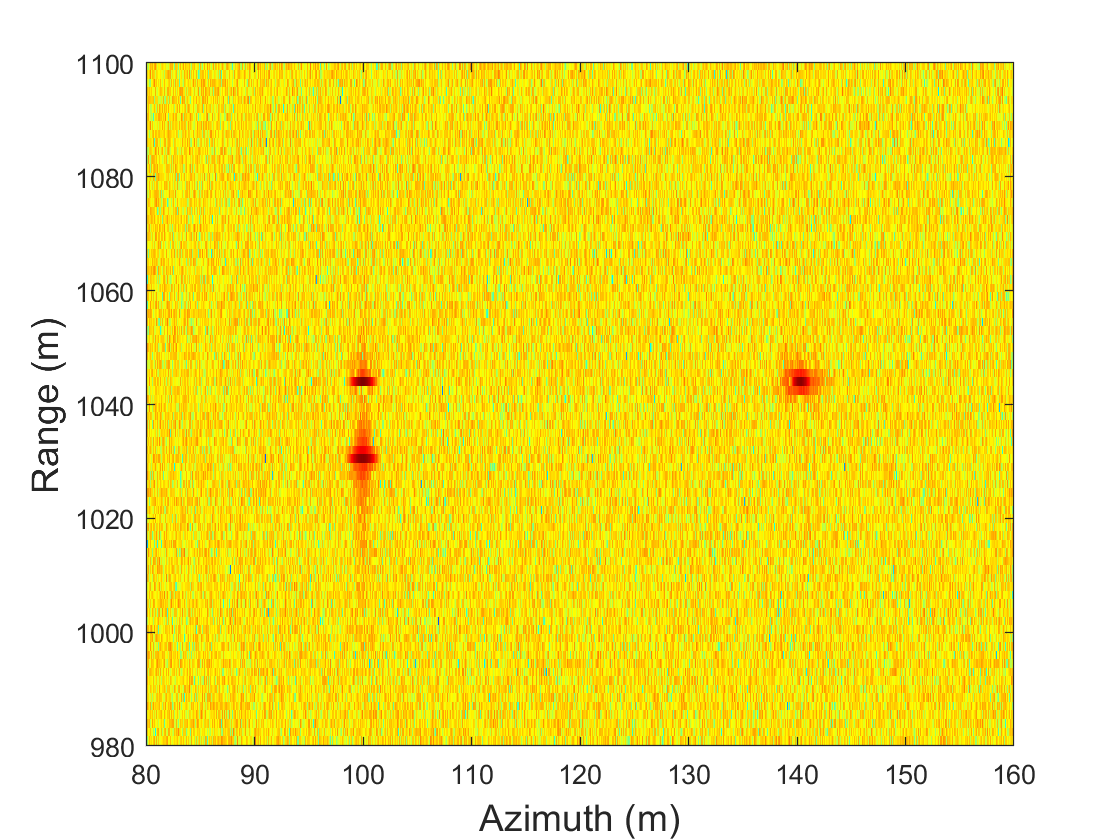}}
		\caption{OFDM-SAR imaging performance: a case of WF and $\mathrm{SNR_{in}}=-20$ dB.}\label{figure3}
	\end{figure*}

	\subsection{Flowchart of OFDM-SAR Imaging}
	First, we validate the flowchart of OFDM-SAR imaging by reconstructing three point targets with normalized RCS values: 
	\begin{itemize}
		\item Target 1: $x_1=300$ m, $y_1=100$ m, and $\bar{R}_1=1044$ m.
		\item Target 2: $x_2=250$ m, $y_2=100$ m, and $\bar{R}_2=1031$ m.
		\item Target 3: $x_3=300$ m, $y_3=140$ m, and $\bar{R}_3=1044$ m.
	\end{itemize}
	
	In this subsection, we present the processing flow of the OFDM-SAR imaging procedure. To illustrate the key steps in a representative scenario, we fix the input SNR at $\mathrm{SNR_{in}} = -20$ dB and adopt the WF-based scheme as the TF filter. A detailed performance comparison among different TF-domain filtering strategies under varying $\text{SNR}_\text{in}$ levels is deferred to the next subsection.
	As shown in Fig.~\ref{figure3}(a), three RCM trajectories exist. To map them onto a common reference scale, the SAR data are first transformed into the RD domain, where the two trajectories corresponding to identical ranges but different azimuth positions become aligned, as illustrated in Fig.~\ref{figure3}(b). Subsequently, the RCMC in Fig.~\ref{figure3}(c) further flattens these two trajectories along the azimuth dimension. Finally, the azimuth compression produces the high-resolution point-target reconstruction shown in Fig.~\ref{figure3}(d), whose focused responses closely match the true target locations.

	\begin{figure*}[!t]
		\centering 
		\subfigure[$\text{SNR}_\text{in}$=-20 dB]{
			\includegraphics[width=2.25in]{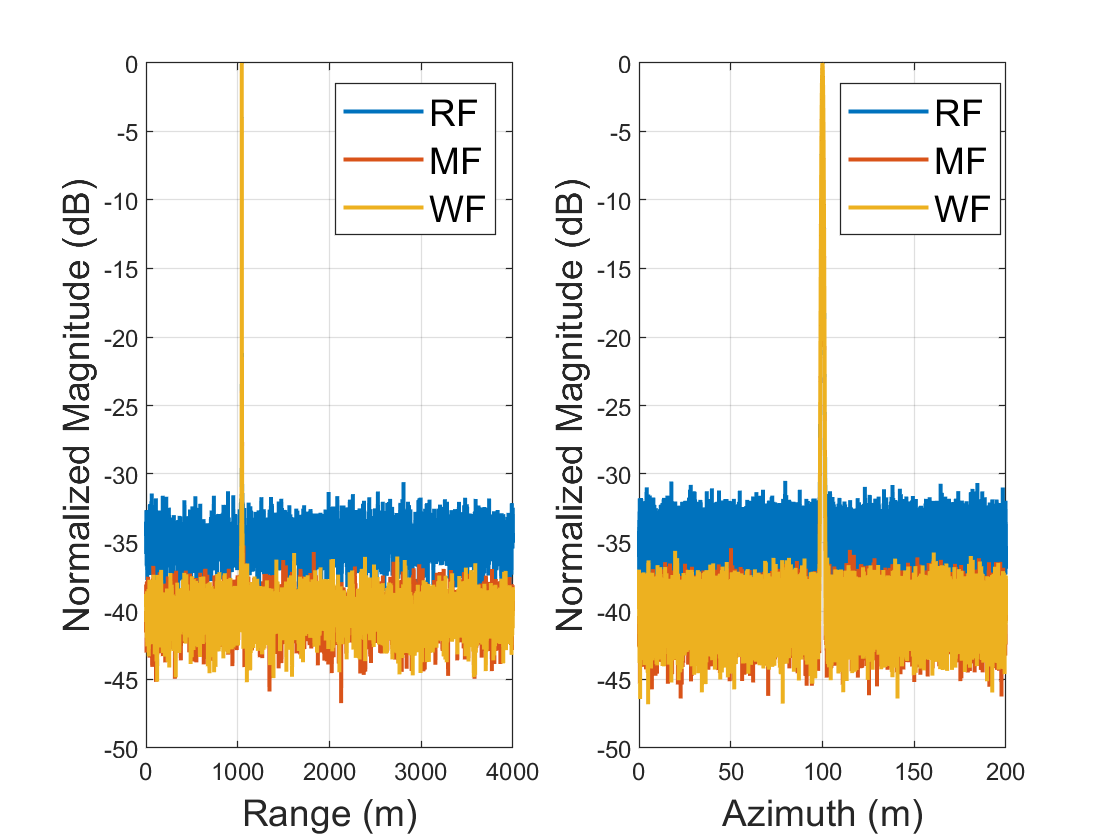}}
		\subfigure[$\text{SNR}_\text{in}$=5 dB]{
			\includegraphics[width=2.25in]{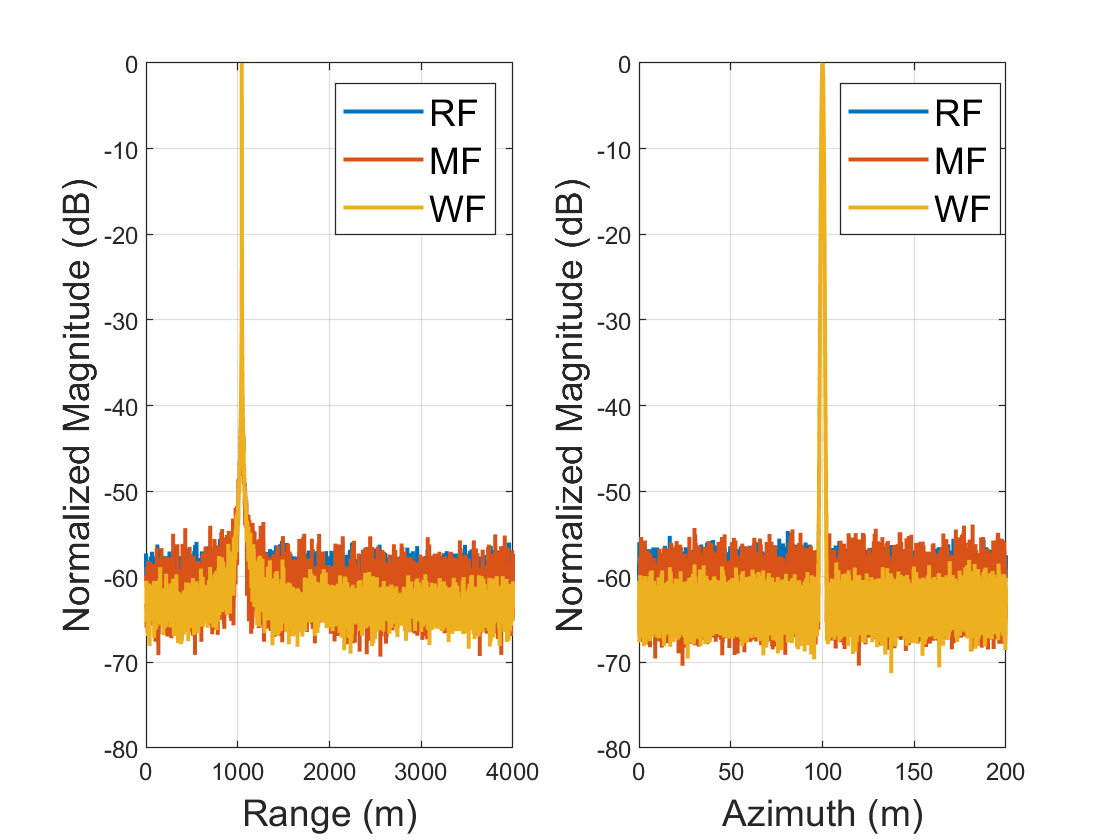}}
		\subfigure[$\text{SNR}_\text{in}$=20 dB]{
			\includegraphics[width=2.25in]{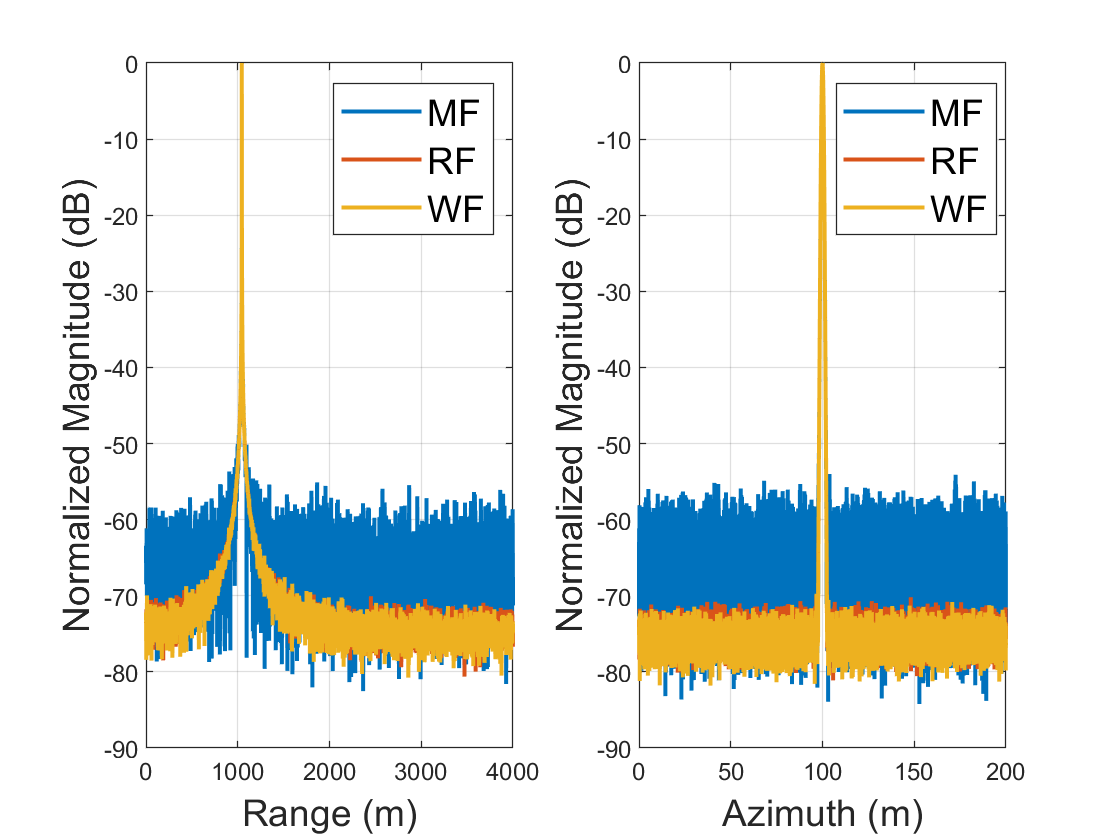}}
		\caption{Range and azimuth results of SAR imaging for different TF-domain filtering schemes under different input SNR values.}\label{figure2}
	\end{figure*}

	\subsection{Comparison Among Different TF Filters}	
	We now focus on the comparison among different TF-domain filtering schemes. To this end, we consider ``Target 1'' as the imaging object and present its separated range and azimuth profiles in Fig.~\ref{figure2} for RF, MF, and WF under $\text{SNR}_\text{in}$ levels of -20 dB, 5 dB, and 20 dB, respectively.
	
	In the low $\text{SNR}_\text{in}$ regime, the filtering performance is predominantly determined by the output noise power, and MF and WF perform significantly better than RF. This is because, with RF, the element-wise division applied to non-constant modulus symbols leads to a substantial amplification of the output noise. In the medium $\text{SNR}_\text{in}$ regime, the sensing performance is jointly affected by the signaling randomness and the amplified noise power, while WF still provides the best focusing performance. In the high $\text{SNR}_\text{in}$ regime, RF and WF achieve similar sensing performance and clearly outperform MF, as the impact of randomness becomes dominant over noise and can be effectively mitigated via the reciprocal operation. Overall, WF outperforms both RF and MF across all considered $\text{SNR}_\text{in}$ conditions\footnote{The transmitted power of conventional spaceborne and airborne SAR systems is typically high. In contrast, low-power, miniaturized UAV-SAR systems generally has a much lower power, especially when the communication module is multiplexed for imaging purposes. From this perspective, MF/WF may be a more suitable choice for the considered LAWNs scenario.}.

	\begin{figure}[!t]
		\centering
		\includegraphics[width=3.0in]{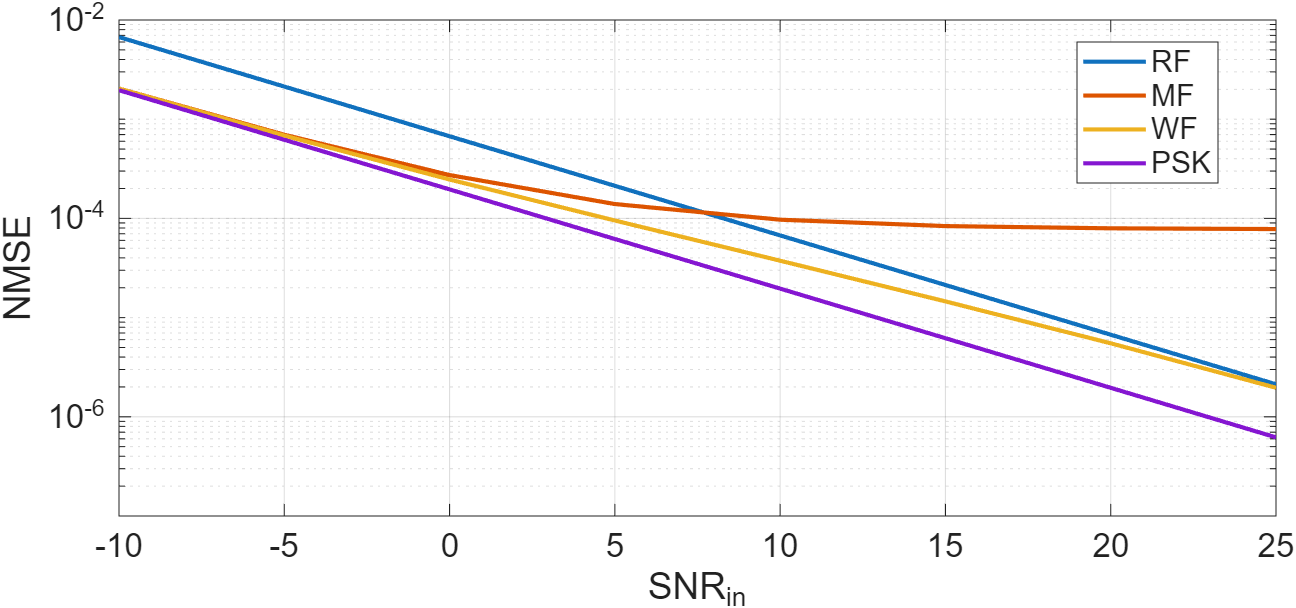}
		\caption{NMSE versus $\text{SNR}_\text{in}$ with different TF-domain filtering schemes.}
		\label{figure6}
	\end{figure}

	To more precisely quantify the imaging performance among RF, MF and WF, we plot the NMSE versus $\text{SNR}_\text{in}$ in Fig.~\ref{figure6}, using the phase shift keying (PSK)-modulated data symbols as the sensing baseline. As expected, WF becomes equivalent to RF in high $\text{SNR}_\text{in}$ regime ($\mathrm{SNR_{in}\geq 25 \ dB}$) and approaches MF in low $\text{SNR}_\text{in}$ regime ($\mathrm{SNR_{in}\leq -5 \ dB}$). These trends are consistent with the qualitative observations in Fig.~\ref{figure2}.

	\subsection{Comparison Among Pilot-only and Data-aided Schemes}
	We now explore the potential of data-aided imaging and illustrate its superiority over the pilot-only scheme. Unless otherwise specified, MF is employed under $\text{SNR}_\text{in}$=5 dB in this subsection. As a benchmark, the pilot-only scheme relies on SRS, which is mapped onto the OFDM subcarriers using a comb-pattern, resulting in a sparse, comb-shaped frequency spectrum. Referring to the 5G NR uplink frame structure \cite{etsi38138}, the SRS occupies $24$ resource blocks, where each block contains $12$ consecutive subcarriers. Therefore, the total occupied bandwidth of the SRS is $24\times12\times\Delta f=8.64$ MHz, which is less than one-tenth of $B_r$. In the simulation, the SRS subcarrier indices range from 
	$1667$ to $1954$. It is worth noting that the comb-pattern is repeated with a frequency spacing of 4 subcarriers. Additionally, SRS can be transmitted with various periodicities. Below we consider two SRS configurations, i.e., a long periodicity of $20$ slots and a short periodicity of $2$ slots, where each slot contains $14$ symbols.

	In contrast, one may also exploit all data symbols for sensing purposes, although this may lead to a substantial computational burden. According to the adopted system parameters, there are $M=T_a/T_\mathrm{sym}=48005$ symbols, corresponding to a PRF of $\frac{1}{T_\mathrm{sym}}\approx 24$ kHz, which is significantly larger than twice the azimuth bandwidth, i.e.,
	$2B_a=2K_aT_a \approx 224$ Hz. To reduce the complexity of our validation, we select one symbol out of every ten for imaging, i.e., operating at one-tenth of the original PRF. Although this down-sampling strategy reduces the energy accumulation and thereby degrades the $\mathrm{SNR}_\mathrm{out}$, the azimuth resolution actually remains unchanged because the synthetic aperture time $T_a$ is fixed.

	
	\begin{figure}[!t]
		\centering
		\includegraphics[width=3.0in]{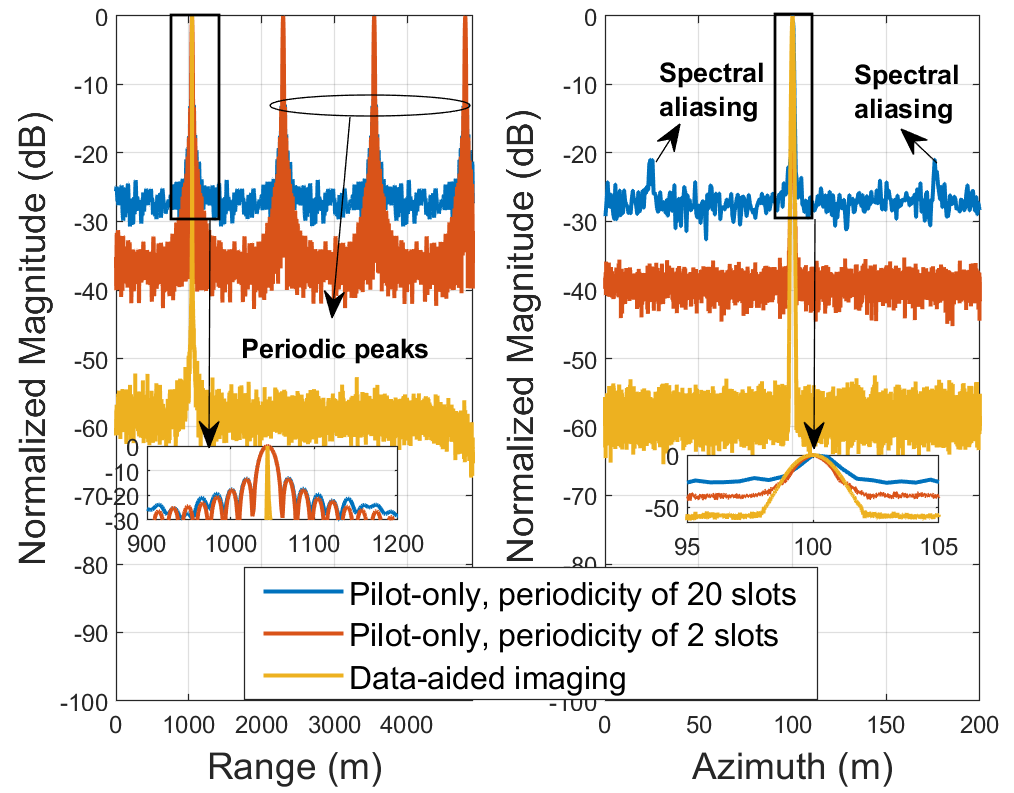}
		\caption{SAR imaging of a reference point target with pilot-only imaging, and data-aided imaging schemes, respectively.}
		\label{figure4}
	\end{figure}	
	
	\begin{figure}[!t]
		\centering
		\includegraphics[width=3.0in]{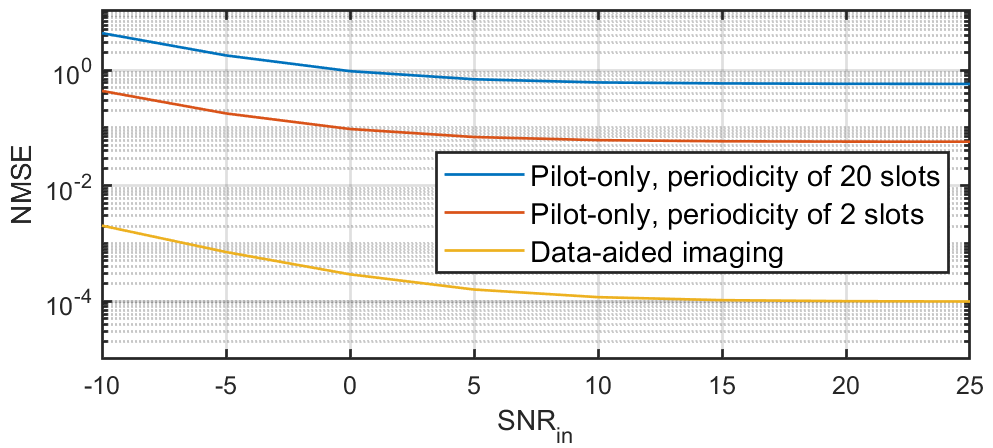}
		\caption{NMSE versus $\text{SNR}_\text{in}$ with pilot-only and data-aided imaging schemes.}
		\label{figure10}
	\end{figure}	
	
	As shown in Fig.~\ref{figure4}, the pilot-only imaging scheme suffers from severe defocusing: multiple peaks appear in the entire range-azimuth image, leading to significant distortion in the response of reference point target. We attribute this behavior to two reasons. First, in the azimuth domain, the PRF of the SRS is $\frac{1}{20\times 14\times T_\mathrm{sym}} \approx 85.7\ \mathrm{Hz} < 2B_a$, which results in additional azimuth peaks due to spectral aliasing caused by undersampling. It is worth noting that the aliasing peaks (with the SRS configuration of $20$ slots' periodicity) are substantially suppressed relative to the target peak, owing to the modulation of the azimuth beampattern, which generally exhibits a $\text{sinc}^2$ shape \cite{cumming2005digital}.
	However, for the SRS configuration of $2$ slots' periodicity, the spectral aliasing in the azimuth can be eliminated due to oversampling. Second, periodic peaks along the range dimension occur because of the discontinuous comb-structure across subcarriers \cite{liu2016space}, where the adjacent peak interval is $\frac{1}{4\Delta f}=8.33 \ \mu s$, corresponding to a round-trip range of 1250 m. 
	In contrast, the data-aided scheme yields a well-focused image, which significantly outperforms the pilot-only benchmark. The involvement of more data symbols enhances the energy accumulation, thereby providing a higher SNR gain. Additionally, Fig.~\ref{figure4} confirms the identical azimuth resolution of three schemes due to the fixed $T_a$, while the range resolution can be significantly facilitated with data-aided scheme, as it occupies the entire $B_r$.

	Furthermore, as depicted in Fig.~\ref{figure10}, the NMSE of data-aided scheme outperforms that of pilot-only schemes by several orders of magnitude, highlighting the benefit of employing more data symbols for sensing. Notably, the relatively large NMSE of the pilot-only schemes arises not only from the reduced number of effective symbols, but also from pseudo and aliasing peaks. Moreover, the NMSE is not directly tied to resolution. Therefore, it is suggested to jointly examine the NMSE of reference point target and its resolution within the region of interest, to calibrate the imaging performance.

\begin{figure*}[!t]
		\centering 
		\subfigure[Reference SAR image \cite{wang2019sar}]{
			\includegraphics[width=1.65in]{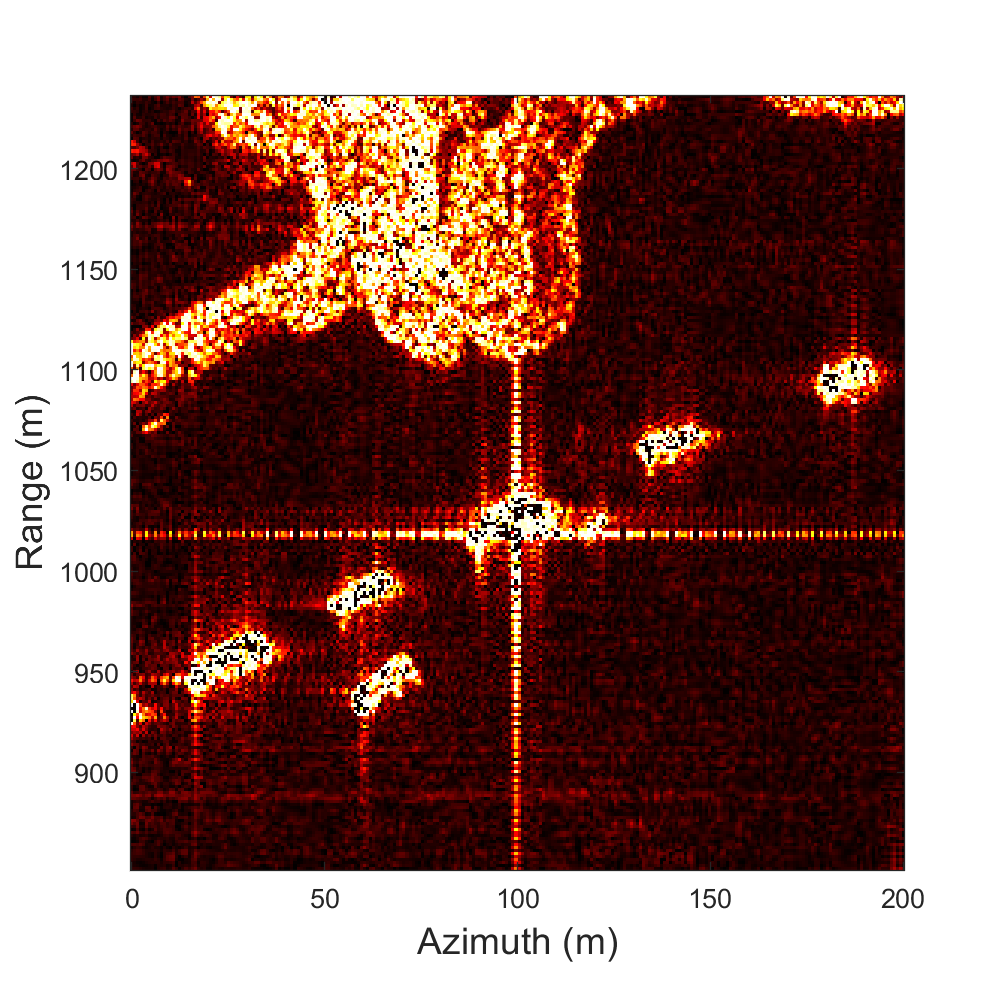}}
		\subfigure[Pilot-only: periodicity of 20 slots]{
			\includegraphics[width=1.65in]{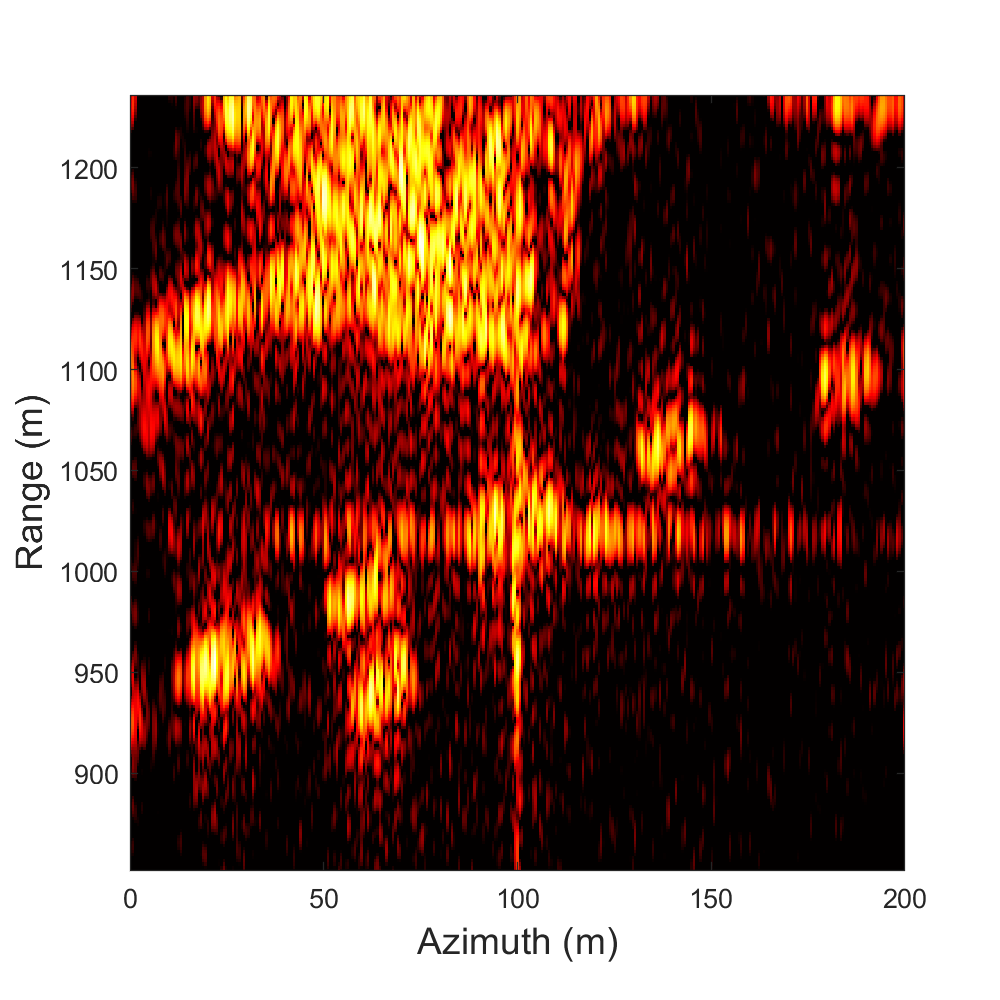}}
		\textit{\subfigure[Pilot-only: periodicity of 2 slots]{
				\includegraphics[width=1.65in]{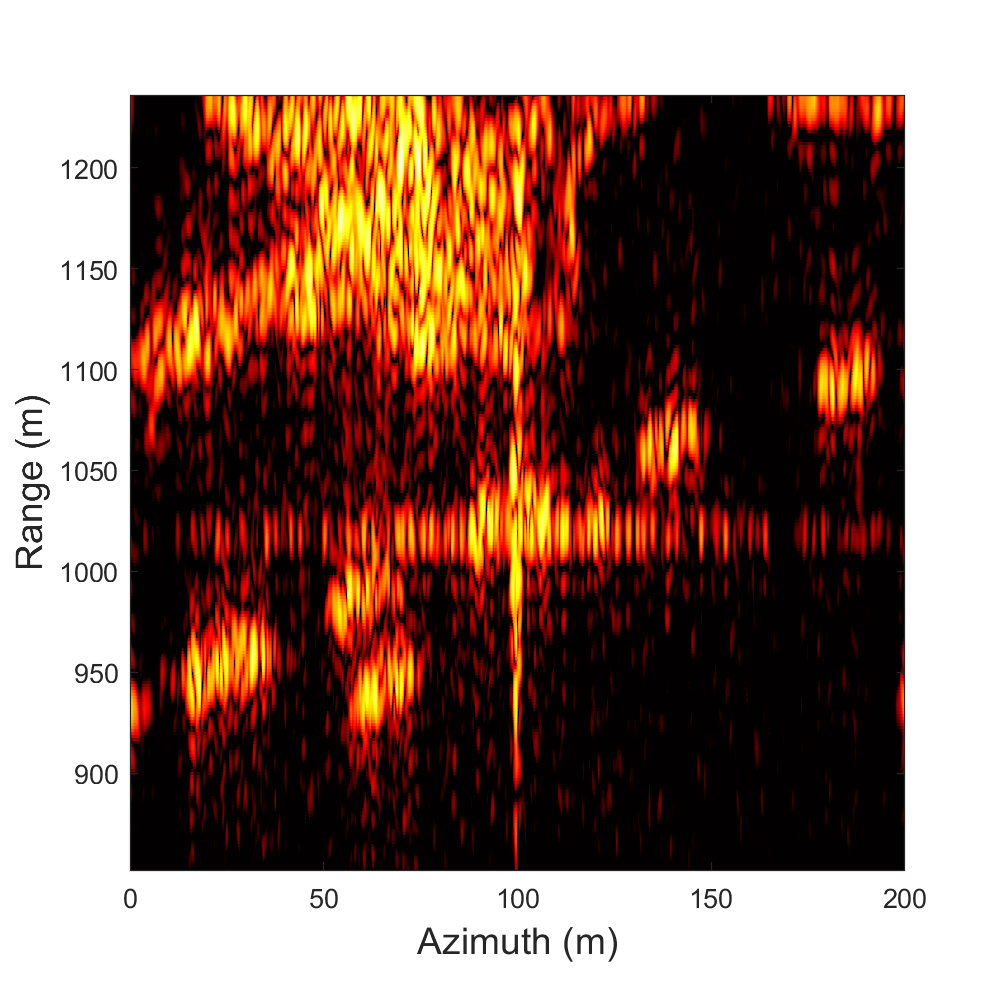}}	}
		\subfigure[Data-aided imaging]{
			\includegraphics[width=1.65in]{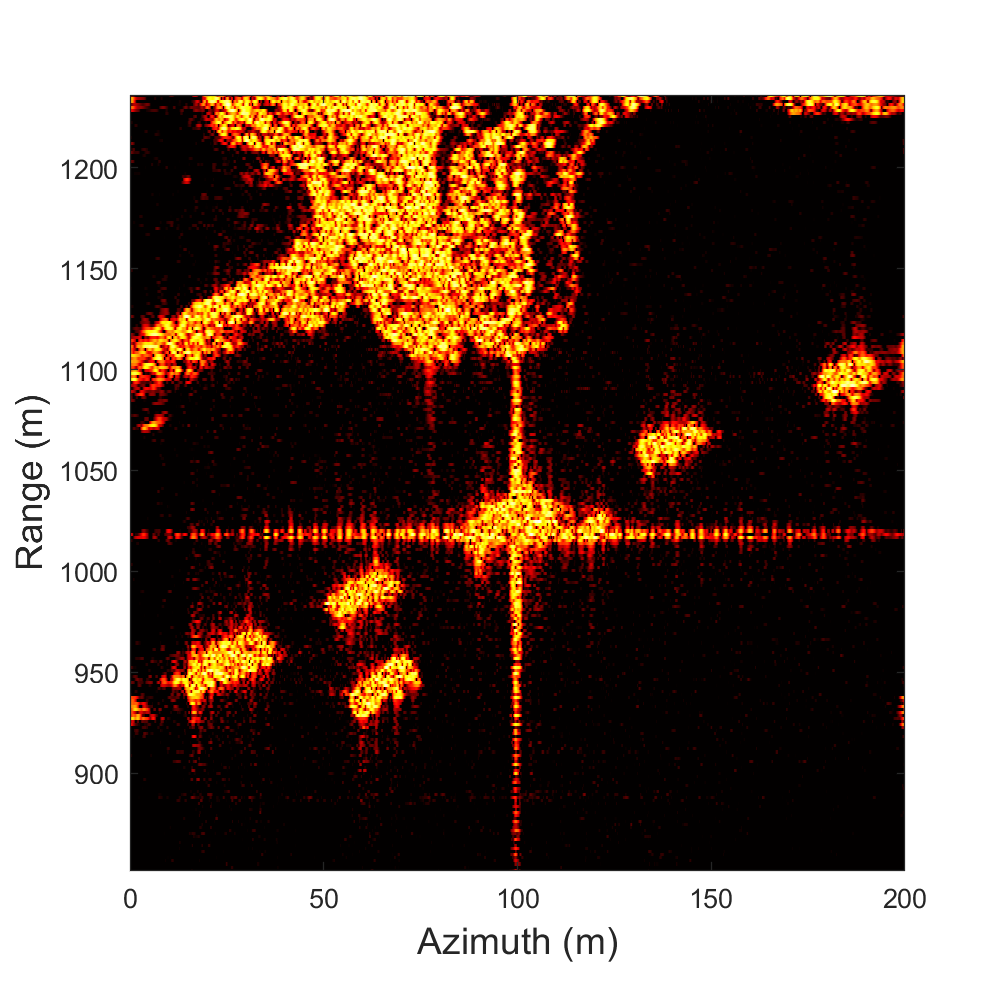}}
		\caption{Reference image of an extended scene, and its reconstructed SAR images with pilot-only and data-aided imaging schemes.}\label{figure5}
	\end{figure*}

	In Fig.~\ref{figure5}, we further reconstruct the SAR image of an extended scene, where a real SAR image \cite{wang2019sar} in Fig.~\ref{figure5}(a) as the reference image for SAR data reconstruction. Evidently, the imaging performance with pilot-only schemes is unsatisfactory, as depicted in Fig.~\ref{figure5}(b) and Fig.~\ref{figure5}(c). This is because SRS occupies only a small fraction of the available bandwidth, leading to limited achievable range resolution. Additionally, the fewer symbols involved leads to a modest SNR gain. Consequently, the reconstructed scene suffers from noticeable blurring and distortion. In contrast, the data-aided scheme fully exploits the entire bandwidth and a denser set of symbols over the synthetic aperture, which jointly enhance both the effective range resolution and the SNR, without additional pseudo peaks. As a result, the point scatterers are much better focused and more faithfully reconstructed, demonstrating a pronounced performance improvement over the pilot-only imaging scheme in practical SAR scenes.

	\section{Conclusion}\label{sec6}
	In this paper, we presented a low-altitude SAR imaging framework that reuses native 5G NR OFDM waveforms and extends conventional pilot-based SAR to fully collect data symbols for sensing. To cope with the performance loss induced by discrete QAM data symbols, we embedded several tailored TF-domain filtering schemes into the conventional RD processing chain, which reveals the mapping mechanism between the signaling randomness and the imaging quality. We further introduced the NMSE of a reference point-target's profile as a complementary metric that jointly reflects $\text{SNR}_\text{out}$, ISLR, and PEL. Simulation results confirm that the proposed data-aided imaging approach substantially outperforms the pilot-only baseline, highlighting its strong potential for integrating low-altitude SAR imaging into UAV data backhaul.

	\appendices
	
	\section{Proof of Lemma 1}\label{appendix1}	
	We may exploit the correlation function of a given signal to characterize how fast it varies by introducing the notion of correlation time. Specifically, the correlation function of 
	$r_m\left(k-k_{q,m}\right)$ is derived in (\ref{equ53}).
	
	\begin{figure*}[!t]			
		\rule{18cm}{1.0pt}
		\begin{equation}
			\begin{aligned}\label{equ53}
				C(\Delta m) & = \mathbb{E}\{r_m\left(k-k_{q,m}\right)r^*_m\left(k-k_{q,m+\Delta m}\right)\}
				= \mathbb{E}\left\{\sum\nolimits_{n,n'}\chi_{n,m}\chi_{n',m+\Delta m}e^{j\frac{2\pi n}{N}\left(k-k_{q,m}\right)}e^{-j\frac{2\pi n'}{N}\left(k-k_{q,m+\Delta m}\right)}\right\}
				\\ & = \sum\nolimits_{n}\mathbb{E}\left\{\chi^2_{n,m}\right\} + \sum\nolimits_{n,n'}\mathbb{E}\left\{\chi_{n,m}\right\}\mathbb{E}\left\{\chi_{n',m+\Delta m}\right\}e^{j\frac{2\pi n}{N}\left(k-k_{q,m}\right)} e^{-j\frac{2\pi n'}{N}\left(k-k_{q,m+\Delta m}\right)} - \sum\nolimits_{n}\mathbb{E}^2\left\{\chi_{n,m}\right\}
				\\ & = N^2 \mathbb{E}^2\left\{\chi\right\}\text{sinc}\left(k-k_{q,m} \right)\text{sinc}\left(k-k_{q,m+\Delta m}\right) + N\mathrm{Var}(\chi)
			\end{aligned}
		\end{equation}
		\rule{18cm}{1.0pt}
	\end{figure*}	
	
	As implied by $k_{q,m}=(\Delta R_{q,m}+\bar{R}_q)/\rho_r$, the correlation time of envelope $r_m\left(k-k_{q,m}\right)$ can thus be inferred by letting $k_{q,m+\Delta m} - k_{q,m} = 1$,
	or equivalently $\Delta R_{q,\Delta m} = \rho_r$.
	This yields the approximated correlation time of envelope as
	\begin{equation}
		\begin{aligned} 
			\Delta m_\text{envelop} \approx \frac{\sqrt{2\rho_r\bar{R}_q}+y_q}{vT_{\mathrm{sym}}}.
		\end{aligned}
	\end{equation}
	
	Similarly, the correlation time of a quadratic/chirp-like azimuth phase term $e^{-j\frac{4\pi}{c} f_c\Delta R_{q,m}}$, can be approximated as
	\begin{equation}
		\begin{aligned} 
			\Delta m_\text{phase} \approx \frac{\lambda R_q}{2\pi Mv^2T^2_\text{sym}}.
		\end{aligned}
	\end{equation}
	
	For a typical low-altitude SAR scenario at the sub-6 GHz band, the synthetic aperture length is on the order of a hundred meters, whereas the detection range can reach the kilometer scale. Consequently, it is readily verified that
	\begin{equation}
		\begin{aligned} 
			\frac{\Delta m_\text{envelop}}{\Delta m_\text{phase}} = \frac{2\pi MvT_{\mathrm{sym}}\left(\sqrt{2\rho_r\bar{R}_q}+y_q\right)}{\lambda R_q} \gg 1,
		\end{aligned}
	\end{equation}
	which consequently proves Lemma 1.

	\section{Proof of Theorem 1}\label{appendix2}
	The stationary point can be solved by letting 
	\begin{equation}
		\begin{aligned}
			\Phi'(\tilde{m}) = \phi'(\tilde{m})-2\pi p/M + \frac{d}{dm}\arg w_m|_{\tilde{m}} = 0,
		\end{aligned}
	\end{equation}
	where the third terms is nearly zero as the envelop varies slowly.
	This results in
	\begin{equation}
		\begin{aligned}
			\phi'(\tilde{m}) =2a\tilde{m}+b \approx 2\pi p/M,
		\end{aligned}
	\end{equation}
	and thus yields (\ref{posp}).
	
	Next, we let $\Phi'(m)$ be expanded at the stationary point with the second-order Taylor expansion as
	\begin{equation}
		\begin{aligned}
			\Phi(m) & \approx \Phi(\tilde{m}) +\Phi'(\tilde{m})(m-\tilde{m}) +\frac{1}{2}\Phi''(\tilde{m})(m-\tilde{m})^2 \\ & = \Phi(\tilde{m}) +\frac{1}{2}\Phi''(\tilde{m})(m-\tilde{m})^2.
		\end{aligned}
	\end{equation}
	Then we have
	\begin{equation}
		\begin{aligned}
			\sum\nolimits_m |w_m| e^{j\Phi(m)}
			= & \sum\nolimits_m |w_m| e^{j\Phi(\tilde{m})} e^{j\frac{1}{2}\Phi''(\tilde{m})(m-\tilde{m})^2}
			\\ \approx & |w_{\tilde{m}}| e^{j\Phi(\tilde{m})} \sum\nolimits_m  e^{j\frac{1}{2}\Phi''(\tilde{m})(m-\tilde{m})^2},
		\end{aligned}
	\end{equation}
	where $\sum_m  e^{j\frac{1}{2}\Phi''(\tilde{m})(m-\tilde{m})^2} \approx \sqrt{\frac{2\pi}{\left\vert\Phi''(\tilde{m})\right\vert}}e^{j \mathrm{sgn}\left(\Phi''(\tilde{m})\right)\frac{\pi}{4}}$ \cite{cumming2005digital},
	which completes the proof.

\ifCLASSOPTIONcaptionsoff
\newpage
\fi
\bibliographystyle{IEEEtran}
\bibliography{reference}

\end{document}